\documentclass{aa}
\usepackage[dvips]{graphicx}
\usepackage[]{amsmath}
\newcommand {\chem}[2] {$\rm{}^{#2}\kern-0.8pt#1$}
\begin{document}

\title{The Modelling of Intermediate-Age Stellar Populations: \\
I- Near-infrared Properties.}
\author{M.\,Mouhcine\thanks{present address: Departement of Physics 
           \& Astronomy, UCLA, Math-Sciences Building 8979, Los Angeles, 
	   CA 90095-1562}, A.\,Lan\c{c}on}
\institute{Observatoire Astronomique, Universit\'e L.\,Pasteur \& CNRS: 
UMR 7550, 11 rue de l'Universit\'e, F-67000 Strasbourg}
\date{Received ; accepted ..}
\titlerunning{Near-IR Models of Intermediate-Age Populations}
\authorrunning{Mouhcine \& Lan\c{c}on}

\abstract{
Evolutionary population synthesis predictions for stellar systems with 
complex star formation histories rest on their major building blocks: 
single-burst population models. In this paper, we discuss how the integrated 
properties of intermediate-age single-burst populations, especially 
in the near-infrared, behave as a function of age and metallicity.\\
Our models take into account all stellar evolutionary phases that 
affect the evolution of the integrated optical and near-infrared 
spectrum of such a population. 
Particular care was dedicated to the Asymptotic Giant 
Branch (AGB) stars, which can be dominant at near-infrared wavelengths. 
First, we present a new synthetic model that
takes into account the relevant physical processes that control the 
evolution through the thermally pulsing AGB, namely (i) the mass-loss, 
(ii) the third dredge-up, and (iii) the envelope burning.
We use this model to evaluate the AGB-termination luminosity, carbon 
star properties as function of initial metallicity and initial mass, 
and the contribution of these stars to the integrated light.
In the isochrones presented in this 
paper the lifetime and the nature of the AGB stars (oxygen-rich or 
carbon-rich) are established as consequences of the interplay 
between the physical processes that control the AGB star evolution. 
The contribution of these stars to the integrated light of the population 
is thus obtained in a consistent way. We optimize our models by using 
a new stellar spectral library that explicitly takes into account 
the spectral features that characterize only AGB stars in comparison 
to other cool and luminous stars. We analyze the contribution 
of the upper AGB to the bolometric and the near-infrared light. Our 
models reproduce the contributions of luminous AGB stars to the bolometric 
and K-band light, and the carbon star contribution to the bolometric light
as observed in the Magellanic Cloud star clusters in a satisfactory 
way, without ad hoc correction factors that could force agreement. \\
Second, we describe the changes occurring in the integrated colours 
when AGB stars first appear. We confirm that, in contrast with the 
classical point of view, no sharp optical/near-infrared colour jump 
occurs when AGB stars start to dominate the stellar population. 
The envelope burning process that affects massive AGB 
stars, making them overluminous with respect to early standard core 
mass-luminosity relations, causes a smoothing of the colour evolution 
for stellar systems dominated by those stars. We reanalyze the 
observational strategy proposed by Lan\c{c}on et al. (1999) to identify 
intermediate-age stellar populations in post-starburst spectra using 
our new model sets. \\
The new spectrophotometric models constitute a first step in a more 
extended study aimed at modelling the spectral properties of the 
galaxies in the near-infrared. 
}
\maketitle
%--------------------------
\section{Introduction}

Evolutionary population synthesis is now a standard tool to investigate 
the evolution of galaxies in the nearby and distant universe. The observed 
features of a galaxy, where a mixture of stellar populations of different 
ages and chemical compositions is present, is modelled by summing spectra 
of appropriate single-burst populations, which themselves are modelled 
by summing the contributions of individual stars of the appropriate 
single age and metallicity (Tinsley 1980, Charlot \& Bruzual 1991, 
Fioc \& Rocca-Volmerange 1997, Maraston 1998).  

The accuracy of the theoretical predictions of population synthesis models 
to study the spectrophotometric evolution of galaxies depends on the quality 
and the completeness of the stellar input physics  
used to model the set of chemically homogeneous aging
single-burst populations. Before facing the problem of modelling 
observed stellar populations with complex star formation histories, the 
reliability of the burst population properties as a function of their two 
fundamental parameters,  age and metallicity, needs to be secured. 
Comparisons of the stellar tracks and derived isochrones with the observed 
features of stellar clusters, for which the internal spread of stellar ages
and metallicities is generally small enough to be consistent with the 
definition of a burst population, is a necessary preliminary task. 

Much effort has been made in recent years to compare 
properties of synthetic burst populations to the observed features
of star clusters (e.g. Arimoto \& Bica 1989, Bruzual \& Charlot 1993, 
Girardi et al. 1995). Very good agreement is now commonly 
found between the observed and the predicted UV and optical properties. 
This expresses our good understanding of 
the main-sequence and early giant phases that dominate UV and optical light. 
Concerning the near-infrared (near-IR), however, 
many problems remain unsolved and 
the agreement between the predicted and observed properties is far from being 
satisfactory. The evolution near-IR colours behaves quite differently from the 
regular trends found in the evolution of UV/visual colours. The near-IR light 
is determined by cool and intrinsically luminous late type stars, especially 
by supergiants (for populations younger than $\sim$\,50\,Myr), AGB stars (for 
populations up to $\sim$\,1.5-2\,Gyr), and Early-AGB (hereafter E-AGB) and 
red giant branch (hereafter RGB) stars (for old 
populations). The contribution of those evolutionary phases to the near-IR 
light varies strongly with age and metallicity, as a consequence of their 
complex evolution that is controlled by the interplay between different
badly-known processes. 

In addition to the difficulties directly related to the
physics of the evolution of the most luminous red stars,
investigations of integrated near-IR properties of intermediate-age 
stellar populations are hampered by large stochastic fluctuations 
in available observations, due to the small number of bright AGB stars 
present in typical star clusters  of Local Group galaxies (Santos \& Frogel 
1997, Lan\c{c}on \& Mouhcine 2000). Until more observations of
very massive clusters become accessible, tests of spectrophotometric 
models remain based on the clusters of the Magellanic Clouds, of
which only a few exceed $10^5$\,M$_{\odot}$. Agreement with the 
trends in these data should of course be sought, but it is clear 
that using them as blind constraints on the evolution of intermediate-age 
stellar populations holds dangers (see also Marigo et al. 1996\,b). 

In the following, we will focus on evolutionary aspects of AGB 
stars that may affect their contribution to the integrated properties 
of a stellar population. Stars with initial masses of 
0.9\,M$_{\odot}\le$M$_{init}\le\,6-8\,$M$_{\odot}$
go through a double-shell thermonuclear burning phase, referred to 
as the Thermally Pulsing AGB (TP-AGB) phase, at the end of their life
on the AGB. The double-shell burning is unstable and leads to thermal
pulses, that drive quasi-periodic luminosity modulations.
In this evolutionary phase, carbon-rich material can be carried up 
to the envelope by a temporary growth of the convective zone, which
is referred to as the third dredge-up.
By mixing additional carbon into the envelope a star can be transformed 
from spectral type K or M (oxygen-rich), to an S-star (C\,$\sim$\,O), 
or to a carbon star (carbon outnumbering oxygen). Spectra of oxygen-rich 
stars are dominated by metal oxide bands such as TiO, VO, and H$_{2}$O, 
whereas carbon stars have bands of C$_{2}$ and CN.
Carbon stars are known to strongly dominate the near-IR light of 
intermediate-age globular clusters (Persson et al.1983, Frogel et al. 1990). 
The efficiency of the formation of carbon stars as function of age and 
metallicity is still matter of debate.

Observations of AGB stars in the Galaxy, LMC, and SMC show that 
the mass-loss rates reached by AGB stars are higher, by orders of magnitude, 
than those typical of RGB stars. The mass loss is the key process 
controlling the lifetime of the TP-AGB. The exact
processes responsible for such high mass-loss rates are still unknown, but 
there is a growing amount of evidence from both observational and theoretical 
points of view, that pulsations coupled with radiation pressure on dust grains 
play a major role. Hydrodynamical models for Long-Period Variable stars (period 
of $100-1000\,$days) show that pulsations may levitate matter to radii where 
grains form efficiently, so that radiation pressure on dust may then accelerate 
the gas beyond the escape velocity (Bowen \& Wilson 1991). Observationally, well 
defined correlations exist between the period of pulsation and the mass-loss 
rates (Schild 1989, Vassiliadis \& Wood 1993, see Habing 1996 for comprehensive 
review).   

For the more massive of the AGB stars, significant burning through the CNO 
cycle can occur at the bottom of the convective envelope, 
a process referred to as envelope burning 
(Bl\"{o}cker \& Sch\"{o}nberner 1991, Boothroyd \& Sackmann 1992, 
Vassiliadis \& Wood 1993). The immediate result is a break-down
of the core mass-luminosity relations long considered "standard", as 
those AGB stars become brighter than otherwise expected (see, e.g.,
Marigo et al. 1999\,b). The higher luminosity triggers higher mass-loss rates,
leading the star to an earlier death. Thus the final core mass attained 
by these stars is lower than it would be if the stars had obeyed
the ``standard" core mass-luminosity relation. This effect can prevent 
an AGB star from becoming a supernova, which has probably had strong 
consequences on the chemical history of galaxies. 
The envelope burning delays the formation of carbon stars, since \chem{C}{12} 
is converted into \chem{C}{13} and \chem{N}{14}, and it gives rise to the 
formation of \chem{C}{13}-rich stars (usually referred to as J-type 
carbon stars) and \chem{N}{14} objects (Richer et al. 1979). 
Theoretical (Boothroyd et al. 1995) and observational (Smith et al. 1995) 
results suggest that the envelope burning 
is a common phenomenon, and that virtually all AGB stars brighter than 
M$_{bol}\,=\,-6$ undergo envelope burning. 

In this paper, we address the theoretical behaviour of the integrated 
colours, focusing on the near-IR properties of intermediate-age 
stellar populations dominated by AGB stars, with the aim of illustrating
and updating our knowledge about the contribution of AGB stars to these
properties. Our further goal is to improve the diagnostic power of 
near-IR spectra of galaxies, to retrieve information about the 
ages and metallicities of their stellar populations.
After a brief statement about background considerations we
consider fundamental (Sect.\,\ref{theoback.sec}),  
Sect.\,\ref{tracks} presents the grid of stellar evolutionary tracks 
constructed to calculate theoretical isochrones. They include all
evolutionary phases from the main-sequence  to the end of the AGB phase. 
Particular attention is paid to the TP-AGB, 
which is included by means of a synthetic model including all
relevant physical processes known to control the evolutionary properties 
that will finally affect integrated stellar population properties. 
In Sect.\,\ref{int_prop} we present the evolution of near-IR 
properties of intermediate-age stellar populations (i.e., contributions 
of AGB stars to the integrated light, broad-band colours, and  narrow-band 
colours). We discuss how they depend on age and metallicity. Comparisons 
with observations in the Magellanic Clouds are shown and discussed. 
Photometric indices that may help detecting intermediate-age populations
in integrated light are re-visited.  
Finally, in Sect.\,\ref{concl} our conclusions are drawn.

%-------------------------------------
\section{Theoretical background}
\label{theoback.sec}

The concept of  {\em fuel consumption} is useful to draw first conclusions 
about the evolution of integrated properties of single-burst populations 
as a function of time, and to become aware of the precautions needed when
constructing self-consistent models. As pointed out by different authors 
(Renzini \& Buzzoni 1986, Girardi \& Bertelli 1998), a relation exists 
between the bolometric luminosity of an evolving burst population 
and the final mass of its current turn-off stars. This is due to two facts:
(i) the lifetime in post-main sequence phases is very short in comparison to 
the main sequence lifetime, and (ii) the luminosity emitted in each 
post-main sequence phase is related to the core mass growth 
during this phase. 
If we consider, in addition, a monotonic relation between the 
initial mass and the total lifetime, 
we can expect that the evolution of the bolometric luminosity of the 
stellar population will just reflect the initial-to-final mass relation. 
Such considerations are also important because the final core mass at 
envelope ejection has direct implications for the 
integrated UV light of old stellar populations (Chiosi et al. 1997).

When modelling the spectrophotometric evolution of stellar populations,
it is fundamental to reproduce the basic observational constraint
of the initial-to-final mass relation before facing other observational 
constraints relative to the integrated properties. 
As intermediate-mass stars lose a large part of their mass during the 
TP-AGB phase, a preliminary task is to model the history of mass loss 
along the AGB phase. 
Detailed modelling of the near-infrared properties requires, moreover,
an accurate representation of the spectral properties 
of AGB stars, as the latter emit the bulk of their energy 
in this range of the spectrum. The two tasks are related since stars with 
massive final cores are, in general, the coolest and the more luminous 
of the stars present.

In this paper we will focus on the evolutionary aspects of the  
AGB, while the average spectra of oxygen-rich and carbon-rich
TP-AGB stars used in our spectrophotometric calculations are 
discussed in a companion paper (Lan\c{c}on \& Mouhcine, 2002; hereafter
Paper~II).

%----------------------------------------------------------
\section{Evolutionary tracks and AGB evolution}
\label{tracks}

In this section we present the evolutionary tracks used
to compute the integrated properties of intermediate-age stellar populations.

\subsection{Up to the end of Early-AGB}

The stellar evolutionary tracks from the main sequence and up to the 
Early-AGB (E-AGB hereafter) phase used in our calculations are those of
Bressan et al. (1994\,a) 
and Fagotto et al. (1994), for [Z\,=\,0.008, Y\,=\,0.25]
and [Z\,=\,0.02, Y\,=\,0.28]. The reader is referred to those papers for 
more details. We note that a convective overshooting scheme was adopted in 
these calculations, lowering the values of the limiting initial masses 
for building degenerate C-O cores after the exhaustion of central helium. 

\subsection{Synthetic evolution of TP-AGB stars}

Full calculations of the evolution of the structure and appearance
of TP-AGB stars are still computationaly expensive,
and only restricted grids in initial mass and metallicity
can be found in the literature. In this 
context, synthetic evolution models provide an attractive
tool to study evolutionary and chemical properties of TP-AGB stars
(Iben \& Truran 1978; Renzini \& Voli 1981; 
Iben \& Renzini 1983;
Groenewegen \& de Jong 1993; Marigo et al. 1996\,a).
Synthetic models are based on analytical representations of
the results of full numerical calculations; they explicitly 
include the physical processes that most directly affect the
evolution along the TP-AGB, and make it possible to test their 
respective effects.

We follow the TP-AGB evolution in a synthetic way, starting from the last 
E-AGB models from the adopted tracks. The total mass M, 
the core mass M$_{c}$, the effective temperature T$_{\rm eff}$, 
the luminosity L, 
the mass loss rate, and the carbon to oxygen ratio of the 
models are left to evolve according to analytical formulae that
inter-relate these quantities and their time derivatives. The equations 
are based on the prescriptions of Wagenhuber \& Groenewegen 
(1998, hereafter WG98), and on previous work by Groenewegen \& de Jong (1993), 
and Vassiliadis \& Wood (1993). The prescriptions of WG98
are a high accuracy reproduction of the results from extensive
grids of complete evolutionary calculations carried out by 
Wagenhuber (1996) for stars in the range 
$0.8\,\rm{M}_{\odot} \leq\,M\,\leq 7\,\rm{M}_{\odot}$ and
metallicities Z=10$^{-4}$, Z=0.008 and Z=0.02.

\subsubsection{Thermal pulses of AGB stars}

Fundamental relations in synthetic AGB models are the core 
mass\,--\,lumin\-osity relation, which gives the maximum luminosity 
during the quiescent hydrogen burning episodes,
and the core mass\,--\,inter\-pulse period relation, which
gives the time separation between two successive thermal pulses 
(Paczy\'nski 1970, Iben \& Truran 1978, Boothroyd \& Sackmann 1988a, 
Bl\"ocker 1995). Predictions of TP-AGB evolution 
are significantly influenced by these input prescriptions: the luminosity
affects the mass-loss rates, and the interpulse period affects the evolution 
of the chemical abundance of the envelope, and hence the spectral properties 
of those stars.

The maximum bolometric luminosity during the quiescent hydrogen burning 
is written as a sum expressing the contributions of different processes (WG98):
\setcounter{equation}{0}
\renewcommand{\theequation}{1\alph{equation}}
\begin{eqnarray}
\label{mc_l_1}   		 
L\,&=&\,18160+3980 \log(Z/Z_{\odot})(M_{c}-0.4468)   \\
\label{mc_l_2}   		 
   & & + 10^{2.705+1.649M_{c}}                   \\  
\label{mc_l_3}   		 
   & & \times 10^{0.0237(\alpha-1.447)M_{c,0}^{2}M_{env}^{2}
                (1-exp(-\Delta M_{c}/0.1))}     \\ 
\label{mc_l_4}   		 
   & & -10^{3.529-(M_{c,o}-0.4468)\Delta M_{c}/0.01} 
\end{eqnarray}
\setcounter{equation}{1}
\renewcommand{\theequation}{\arabic{equation}}
Z denotes the initial metallicity, M$_{c,0}$ is the core mass at the 
first thermal pulse, $\Delta$M$_{c}$ is the core mass growth since the 
beginning of the TP-AGB (i.e. $\Delta$M$_{c}$\,=\,M$_{c}-$M$_{c,0}$),
and M$_{env}$ is the current envelope mass. 
Masses and luminosities are expressed in solar units. 
This core mass-luminosity relation has the advantage of including 
relevant physical processes that were omitted in early versions. 
The first term expresses the usual linear relation in the full amplitude 
regime of the thermal pulses (Paczy\'nski 1975, Vassiliadis \& Wood 1993); 
the second term provides a correction that becomes significant for high core 
mass stars (M$_{c}\,\ge\,$0.95M$_{\odot}$); the third term accounts for 
the excess of luminosity produced by envelope burning, which drops when 
the envelope mass is reduced; the last term provides a correction for the 
``turn-on'' effects, i.e. the progressive growth of the thermal pulses
early on the TP-AGB. This formulation gives us the opportunity to look 
into the effect of the envelope burning, of the convective mixing length 
$\alpha$ and of the chemical composition on the integrated properties. 
Our {\it{standard model}} will use $\alpha=2$ (Bl\"ocker \& Sch\"onberner 
1991).

The core mass\,--\,interpulse period relation is given by:  
\begin{eqnarray}
\log\tau_{ip}&=&(-3.628+0.1337\log(Z/Z_{\odot}))(M_{c}-1.9454) \nonumber \\
             & & -10^{- 2.080-0.353\log(Z/Z_{\odot})
                      + 0.2(M_{env}+\alpha-1.5)}  \nonumber \\
             & & -10^{-0.626-70.30(M_{c,o}-\log(Z/Z_{\odot}))\Delta M_{\rm c}}
\label{tauip}                       
\end{eqnarray}
Again, the first term expresses the standard relation, a decreasing 
function of the core mass with some dependence on the metallicity. 
The second term accounts for envelope burning, by somewhat reducing 
this time interval. The final term corrects for the progressive turn-on
of the pulses before the asymptotic regime is reached.
The influence of the initial metallicity on the 
interpulse period is strong, in the sense that the interpulse period 
increases with decreasing metallicity.

Between two thermal pulses, the luminosity is not constant. The helium
flashes drive deviations from the luminosity given by the core 
mass\,--\,lumin\-osity relation. Those variations have a strong effect 
on the AGB star populations, in particular on the bright tail of the AGB 
luminosity function (Marigo et al. 1996\,a). We use the prescriptions of 
WG98 to calculate the interpulse light curves.

To obtain the position of a star in the HR diagram and to calculate the 
mass loss rate, an estimate of the stellar effective temperature is needed.
We use the analytical relation presented by Wagenhuber (1996).
This relation is based on stellar models with carbon-to-oxygen number
ratios below one. We use it even when stars become carbon-rich, 
although carbon star atmospheres are expected to be more compact
than those of M type stars (Scholz \& Tsuji 1984, Loidl et al. 2001). 
A complete integration of the envelope and atmosphere 
equations with abundance dependent opacities would
provide more consistent positions in the {\em theoretical} HR diagram, 
but our purpose is to predict {\em observable} properties. 
Fundamental uncertainties related to the definition of the effective 
temperature of AGB stars and its relation to observable properties 
(e.g. the available spectra) are discussed in detail by 
Baschek et al. (1991) and in Paper~II.
These uncertainties are large, and justify an {\em a posteriori}
calibration of the relation between the effective temperature of 
TP-AGB stars and their colours in any case. Therefore, the theoretical 
absolute effective temperature scale of the TP-AGB tracks is not critical.

\subsubsection{Mass loss rates}

One of the most important processes controlling the secular evolution
of AGB stars in the HR diagram and their spectral type is the mass loss 
rate (Bedijn 1987, Bryan et al. 1990, Bl\"ocker 1995, Frost et al. 1998, 
Zijlstra 1999). It affects the lifetime and 
the average luminosity of stars in this phase. Both these 
quantities shape the contribution of the whole phase to the integrated 
light. The larger the mass-loss rate, the shorter is the lifetime of the 
TP-AGB phase, the dimmer are its brightest stars (at wavelengths sensitive 
to circumstellar obscuration), and the lower is the contribution to 
the total light of the population. 
In addition, a high mass loss rate favours the formation of 
carbon stars, since less carbon needs to be dredged up to 
change the carbon-to-oxygen abundance ratio of a less massive envelope.
Unfortunately, mass-loss is still poorly understood. The mass loss 
dependence on the fundamental stellar parameters and the mass loss
efficiency $\eta$ are still far from being definitively determined. 
Unless otherwise stated, we will use the analytical mass loss prescription 
of Bl\"ocker (1995), derived from the hydrodynamical models for Long 
Period Variable stars (LPVs) of Bowen (1988):
\begin{equation}
\dot{M} = 4.83 \,\,10^{-9}\,\eta\,L^{2.7}\,M^{-2.1}\,\dot{M_{R}}
\end{equation}
where $\eta$ is the mass loss efficiency and $\dot{M_{R}}$ is Reimers's 
mass loss rate 
(i.e., $\dot{M_{R}}\,=\,1.27\,10^{-5}\,\eta_{R} M^{-1}L^{1.5}T^{-2}_{eff}$, 
with $\eta_{R}=1$). This formulation of the mass loss is particularly
sensitive to the luminosity. In our {\it{standard model}} a value of 
$\eta=0.1$ will be used as suggested by Groenewegen et al (1994) from 
fitting the luminosity function of carbon stars in the LMC. In stars with 
relatively high initial masses, the strong decrease of the current mass 
along the TP-AGB contributes to producing the observed superwinds, that 
terminate this evolutionary phase. The effect of the effective temperature 
on the evolution of $\dot{M}$ at a given metallicity is smaller. 
The average T$_{\rm eff}$ of the evolutionary tracks however depends on 
metallicity, which contributes to lowering the mass loss of metal deficient 
stars. We do not consider an explicit metallicity dependence in the mass 
loss rate.   

\subsubsection{The rate of evolution}

Hydrogen burning is the dominant source of energy during most of the interpulse 
period. The equation that describes the core growth in M$_{\odot}$ yr$^{-1}$ 
reads:
  \begin{equation}
  \frac{dM_{c}}{dt}=q(Z)\frac{L_{H}}{X}
  \label{dMcdt}
  \end{equation}
where $q(Z)$, the mass burnt per unit energy release is slightly metallicity
dependent, X is the hydrogen abundance by mass in the envelope, and L$_{H}$
the luminosity provided by the burning of hydrogen. Since there is some 
contribution to the total energy from the core shrinking and helium burning,
the luminosity produced by the hydrogen burning that enters Eq. \ref{dMcdt}
is smaller than the total luminosity derived from the core mass-luminosity 
relation, so that (WG98):
  \begin{equation}
  \log\left(\frac{L_{H}}{L}\right) = -0.012 - 10^{-1.25-113 
                      \Delta M_{c}}-0.0016 M_{env}
  \label{Hydr_burn}
  \end{equation}
Here L is is the luminosity derived using the core mass-luminosity 
relation, with the term providing a correction for the envelope burning 
(e.g. Eq.\,\ref{mc_l_3}) set to unity. 

\subsubsection{Chemical evolution of TP-AGB stars}

Between the main sequence turn-off and the tip of the AGB phase, low 
to intermediate-mass stars experience three types of dredge-up episodes, 
during which material that underwent nuclear burning is mixed to the 
surface by envelope convection. During the first dredge-up episode, 
where the star reaches its Hayashi track after the hydrogen exhaustion 
in the core, and the second dredge-up at the base of the E-AGB, 
products from the CNO cycle are transported to the surface; third 
dredge-up episodes during the TP-AGB bring up helium shell burning 
products as well. In our calculations, the chemical abundances 
prior to the onset of the third dredge-up during the TP-AGB phase 
are directly taken from the evolutionary tracks used for the 
evolutionary phases earlier than the TP-AGB phase.

The third dredge-up plays a crucial role in the formation of carbon 
stars as it leads to the enrichment of the surface in carbon. 
As most convection-determined phenomena, dredge-up is not yet
fully understood. The amount of \chem{C}{12} enhancement increases 
strongly with the metallicity and mass (Wood 1981, Lattanzio 1987, 
Boothroyd \& Sackmann 1988b, Vassiliadis \& Wood 1993). 
Most stellar evolution calculations fail to predict carbon 
stars at luminosities as low as observed,  and are only able 
to obtain dredge-up for  relatively high initial stellar masses 
(M$\,\ge\,1.5\,$M$_{\odot}$; e.g. Bl\"ocker 1995, 
Forestini \& Charbonnel 1997, Wagenhuber \& Weiss 1994). Boothroyd \& Sackmann 
(1988b) were able to find dredge-up in a 0.81 M$_{\odot}$ and Z$=10^{-3}$ star, 
but using a very high mixing length parameter (i.e. $\alpha\,=\,3$), which is 
questionable. There is still much debate whether or not material is dredged up 
at every thermal pulse, and how much (Bl\"ocker 1999, Mowlavi 1999).
The observed carbon stars luminosity function requires that the dredge-up 
occurs efficiently to produce carbon stars for initial masses of 
about $\sim\,1.3-\,1.4\,$M$_{\odot}$ (Groenewegen et al. 1995, 
Marigo et al. 1996\,a). 

The semi-analytical treatment of the third dredge-up process used in 
this paper requires three basic inputs:
(i) the critical core mass M$^{min}_{c}$ above which it is assumed that
dredge-up occurs (we relax the condition that the thermal pulse 
must have reached the full amplitude regime, introduced as an
assumption by Groenewegen \& de Jong 1993), 
(ii) the dredge-up efficiency parameter, defined as the fraction of 
the core mass growth that will be dredged up to the envelope 
($\lambda=\Delta M_{dredge}/\Delta M_{c}$, where 
$\Delta M_{c}\,=\,\int_{0}^{\tau_{ip}}\,(dM_{c}/dt)dt$), and (iii) the 
composition of the material in the convective inter-shell after a thermal 
pulse and before the penetration of the convective envelope; this is the 
material that will be dredged-up, mainly consisting of helium, carbon, 
and oxygen. We have adopted $M^{min}_{c}\,=\,0.58$\,M$_{\odot}$ 
and $\lambda\,=\,0.75$, as required to
reproduce the luminosity function of field carbon stars in the LMC
(Groenewegen \& de Jong 1993; see also Marigo et al. 1996\,a, 1999\,a), 
and the chemical composition of the convective inter-shell is taken from 
Boothroyd \& Sackmann (1988a).

The second effect which can heavily alter the surface chemical composition 
of TP-AGB stars is envelope burning. This effect prevents or delays the 
conversion to carbon stars for massive AGB stars, because of the efficient 
burning of carbon-rich dredged-up material to nitrogen. 
The physical conditions needed to trigger this process, and its efficiency, 
are still uncertain. To take envelope burning into account, 
we procede as follows. We assume that envelope burning is effective
(e.g. that CNO cycle is operating at the base of the envelope), for 
sufficiently luminous stars ($\log(L/L_{\odot})\,\ga\,4.2$), only 
if the envelope mass exceeds a certain value M$_{env}^{EB}$ 
(see Marigo, 1998, for an alternative approach). As an estimate of
this critical value, we use the approximation derived by WG98: 
\begin{equation}
 M_{env}^{EB}\,=M(1-0.2\alpha\,-\,5Z)
\end{equation}
where M$_{env}^{EB}$ and M are, respectively, the current envelope mass 
and total mass. $\alpha$ is the mixing length parameter, and $Z$ is 
the stellar initial metallicity. 
Within the framework of our semi-analytical treatment,
dredge-up can occur even after envelope burning has stopped
near the end of the TP-AGB. This is of vital importance to 
reproduce the formation of observed luminous carbon stars 
(van Loon et al. 1998, Frost et al. 1998, Mouhcine \& Lan\c{c}on 2002).

Denoting by X$_{i}^{new}$, X$_{i}^{shell}$ the abundance of the element 
$i$ in the envelope and in the inter-shell, by M$_{env}$ the mass of the 
envelope just before the pulse, and by $\Delta\,$M$_{dup}$ the mass 
dredged up to the envelope after the pulse, the new surface abundance 
after the pulse is written: 
\begin{equation}
X_{i}^{new}=\frac{X_{i}^{old}M_{env}+X_{i}^{shell}\Delta M_{dup}}
                 {M_{env}+\Delta M_{dup}}+\Delta X_{i}^{EB}
\label{chem_evol}
\end{equation}
The second term in the right side of equation \ref{chem_evol} takes into 
account the abundance variation due to envelope burning. To calculate the 
amount of the dredged up material and the envelope material that undergo 
thermonuclear burning at the bottom of the envelope, we follow the 
semi-analytical approach adopted by Groenewegen \& de Jong (1993), 
based on the results of Renzini \& Voli (1981). In order to calculate the 
time evolution of the chemical abundances, we use the approach 
presented by Clayton (1983). The basic assumption in this method is that the 
two reaction chains of the CNO-cycle (i.e., the CN-cycle and the ON-cycle) can 
be separated. The nuclear reaction rates are taken from Fowler et al. (1975).\

\subsection{The end of the AGB evolution}

The TP-AGB phase ends with the total ejection of the 
envelope, when the star moves from the red to the blue side
of the HR diagram. In practice, we follow a star's evolution until the 
envelope mass is reduced below $10\%$ 
of the core mass growth during the last interpulse period (Iben 1985). The 
TP-AGB evolution is also stopped when the core mass reaches the 
Chandrasekhar limit ($\sim\,1.4\,M_{\odot}$) just before carbon deflagration
would occur. In our models however, this condition is never met, 
being prevented under the influence of the stellar mass loss. 
%The time step is determined from the condition that the total mass 
%change by no more value than $10^{-4}\,M_{\odot}$. (AL- Ce n'est qu'une
% des contraintes; les donner soit toutes, soit aucune).

\subsection{Results and constraints}

For the purpose of this paper, we have restricted the computation
of the semi-analytical TP-AGB star evolution to two initial metallicities, 
[Z\,=\,0.008, Y\,=\,0.25] and [Z\,=\,0.02, Y\,=\,0.28].
Each set of tracks covers the initial mass range of  
0.9M$_{\odot}$\,--\,M$_{up}$. M$_{up}$ is taken equal to 5\,M$_{\odot}$ 
for Z=0.008, and equal to 6\,M$_{\odot}$ for Z=0.02. In the following 
subsection, we present and discuss the evolution of 
the properties of TP-AGB stars that are relevant to the purpose of this paper.

Complete grids of evolutionary tracks and isochrones
for an extended metallicity grid (from Z/Z$_{\odot}\,=\,1/50$ 
to Z/Z$_{\odot}\,=\,2.5$) have been computed, and are being used
to interpret carbon star counts in galaxies (Mouhcine \& Lan\c{c}on 2002). 
They will be published elsewhere (in preparation).

\subsubsection{The initial-final mass relation}

\begin{figure}[ht]
\includegraphics[clip=,angle=0,width=9.5cm,height=9.cm]{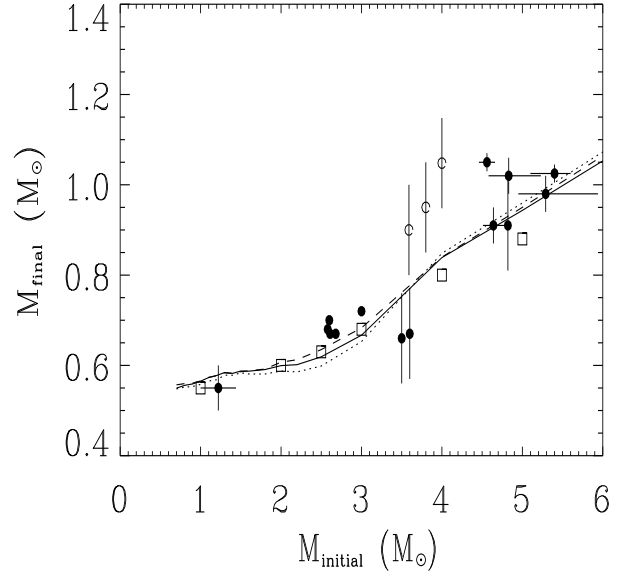}
\caption[]{Final stellar remnant mass after complete ejection of the envelope,
plotted as a function of the initial mass for solar metallicity models,
for three mass-loss prescriptions: Bl\"ocker (1993, continuous line), Reimers
(1975, dotted line), Vassiliadis \& Wood (1993, long-dashed line). 
$\alpha=2$ is assumed. Observational data are taken 
from Jeffries (1997), Herwig (1996), Weidemann (2000, open squares). Open points 
with error bars represent data points of lower quality (cf. Herwig 1996). }
\label{Mi_Mf}
\end{figure}

In Fig.\,\ref{Mi_Mf} the final core mass, M$_{final}$, left after the complete 
ejection of the envelope is plotted against the initial mass M$_{init}$ of each 
model for solar metallicity. The predicted initial-final mass relations for both
metallicities considered in this paper (Z=0.02 and Z=0.008) are tabulated in 
Table\,\ref{lifetimes.tab}, together with the maximum quiescent bolometric 
luminosity reached on the TP-AGB, the total lifetime of the TP-AGB phase, and 
the fraction of this duration spent as a carbon star. The final mass can be 
considered as the mass of the remnant white dwarf.

The theoretical predictions of Fig.\,\ref{Mi_Mf}
are computed using different mass-loss prescriptions, based on those
of Bl\"ocker (1995), but also those of Reimers (1975) and 
Vassiliadis \& Wood (1993). We have modified the mass loss prescriptions 
using calibrated mass-loss efficiencies (Groenewegen \& de Jong 1994, 
Marigo et al. 1996\,a). 
Also shown in the plot is the semi-empirical initial-final mass relation from 
Herwig (1995), Jeffries (1997), and the new revisited relation from Weidemann 
(2000) for the solar neighbourhood.  Very good agreement is seen between the 
predicted solar metallicity initial-final relation and the most recent 
semi-empirical determination. This agreement shows 
that the effects of the envelope burning and the third 
dredge-up are important in determining the evolution of the stellar core mass 
along the TP-AGB (and hence the contribution of AGB stars to the 
integrated light of stellar populations). Indeed, the reduction of the core 
mass by recurrent third dredge-up events keeps the core mass close to 
its value at the begining of the TP-AGB phase. Neglecting these effects, 
as in some previous population synthesis models that take into account 
a simplified TP-AGB phase, leads to the prediction of much higher final 
masses than observed (see for example Bertelli et al. 1994, 
Girardi et al. 2000). The net effect of the choice of one or the other 
mass-loss prescription, in comparison, is small. 

\begin{table}
\centering
\caption{For Z=0.02 and Z=0.008 and for each initial mass 
M$_{init}$ (in M$_{\odot}$), the table lists the values
of the final mass M$_{final}$, the brightest quiescent bolometric 
magnitude M$_{bol}$, the lifetime on the TP-AGB $\tau_{TP-AGB}$ (in Myr), 
and the fraction of the TP-AGB lifetime spent as a carbon star. 
The predicted values are those of our standard model 
($\alpha\,=\,2$, $\eta\,=\,0.1$, and $\lambda\,=\,0.75$)}
\label{lifetimes.tab}

\begin{tabular}{lcccclcccc}
\multicolumn{1}{c}{{Z$_{init}$}}&
\multicolumn{1}{c}{{M$_{init}$}}&
\multicolumn{1}{c}{{M$_{final}$}}&
\multicolumn{1}{c}{{M$_{bol}$}}&
\multicolumn{1}{c}{{$\tau_{TP-AGB}$}}&
\multicolumn{1}{c}{{$\tau_{C}/\tau_{TP-AGB}$}}\\ \hline
\\
0.02   & 0.931  &   0.549  &  -3.952  &  0.175  & 0.000  \\ 
       & 1.000  &   0.556  &  -4.058  &  0.282  & 0.000  \\
       & 1.078  &   0.559  &  -4.161  &  0.423  & 0.000  \\
       & 1.159  &   0.565  &  -4.291  &  0.481  & 0.000  \\
       & 1.242  &   0.573  &  -4.393  &  0.508  & 0.000  \\
       & 1.328  &   0.578  &  -4.469  &  0.588  & 0.000  \\
       & 1.413  &   0.584  &  -4.554  &  0.555  & 0.000  \\
       & 1.499  &   0.582  &  -4.561  &  0.731  & 0.076  \\
       & 1.585  &   0.587  &  -4.594  &  0.791  & 0.083  \\
       & 1.670  &   0.587  &  -4.657  &  0.986  & 0.139  \\
       & 1.840  &   0.591  &  -4.729  &  1.398  & 0.110  \\
       & 2.016  &   0.599  &  -4.799  &  1.591  & 0.329  \\
       & 2.200  &   0.602  &  -4.867  &  2.125  & 0.286  \\
       & 2.500  &   0.619  &  -4.989  &  2.402  & 0.518  \\
       & 3.000  &   0.667  &  -5.331  &  1.503  & 0.455  \\
       & 4.000  &   0.839  &  -6.025  &  1.438  & 0.000  \\
       & 5.000  &   0.943  &  -6.309  &  0.060  & 0.0015 \\
       & 6.000  &   1.053  &  -6.715  &  0.022  & 0.000  \\
       
\hline
0.008  & 0.899  &  0.544   &  -3.942  &  0.365  & 0.000 \\ 
       & 0.949  &  0.549   &  -3.960  &  0.393  & 0.000 \\
       & 1.062  &  0.56    &  -4.175  &  0.518  & 0.000 \\
       & 1.230  &  0.573   &  -4.460  &  0.726  & 0.000 \\
       & 1.404  &  0.581   &  -4.641  &  0.859  & 0.097 \\
       & 1.576  &  0.583   &  -4.717  &  1.065  & 0.158 \\
       & 1.748  &  0.588   &  -4.772  &  1.531  & 0.296 \\      
       & 2.000  &  0.596   &  -4.862  &  2.333  & 0.298 \\       
       & 2.500  &  0.623   &  -5.101  &  2.781  & 0.779 \\
       & 3.000  &  0.685   &  -5.498  &  1.127  & 0.768 \\
       & 4.000  &  0.899   &  -6.22   &  0.0676 & 0.00  \\
       & 5.000  &  0.977   &  -6.432  &  0.0449 & 0.00  \\
\hline
\end{tabular}
%}
\end{table}

\subsubsection{TP-AGB lifetimes}

The contribution of TP-AGB stars to the integrated 
properties of a stellar population depends on the time stars spend on 
the TP-AGB. The lifetime of this phase is mainly determined by the 
mass-loss history, and essentially by the onset of the superwind, 
and it depends in a complex way both on stellar mass and on metallicity 
(Vassiliadis \& Wood 1993). 

\begin{figure}[ht]
\includegraphics[clip=,angle=0,width=9.cm,height=8.cm]{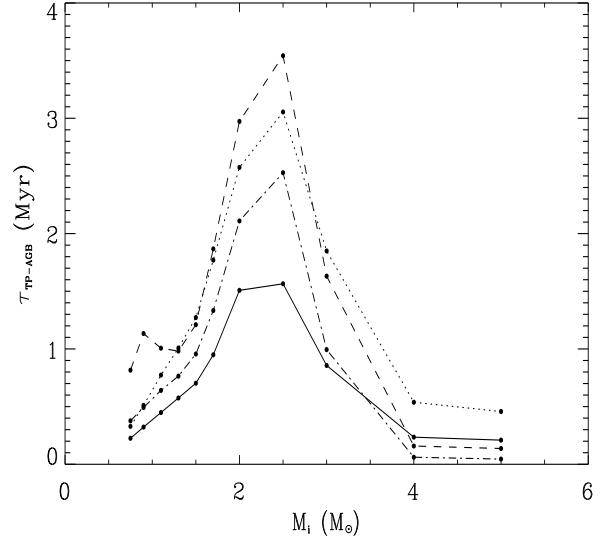}
\caption[]{Theoretical total TP-AGB lifetime as a function of initial mass
for LMC metallicity. Different mass loss prescriptions were used: Reimers' law
(1975; solid line), Bl\"ocker's prescription (1993; dashed), 
Vassiliadis \& Wood (1993; dot-dashed), Baud \& Habing (1983; dotted).}
\label{TPAGBlifetime_dmZlmc}
\end{figure}

\begin{figure}
\includegraphics[clip=,angle=0,width=9.cm,height=8.cm]{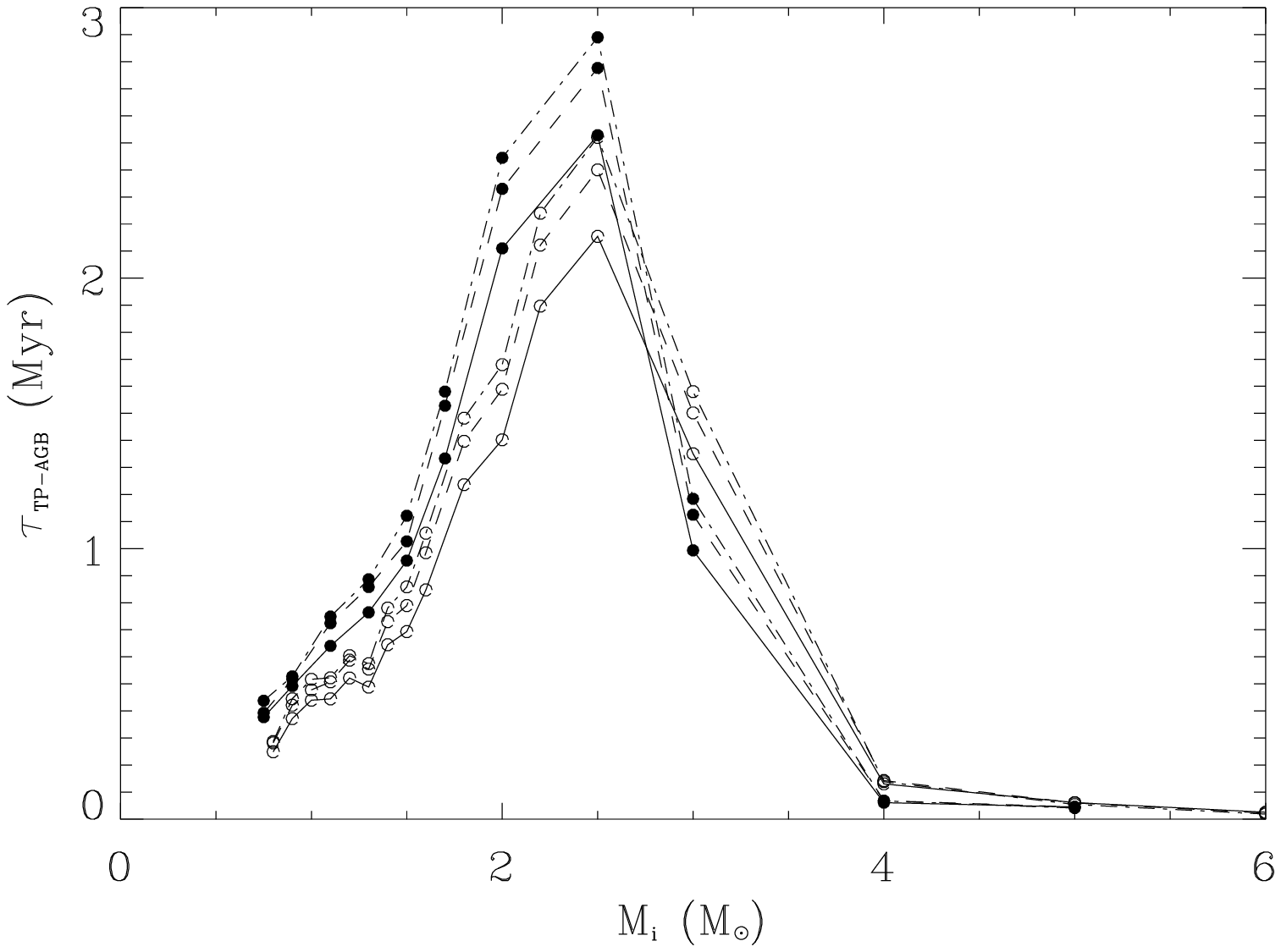}
\caption[]{Total TP-AGB lifetime as a function of initial mass. 
Filled and open circles correspond, respectively, to LMC and solar 
metallicities. Different mixing length
parameters were used in the calculations: $\alpha=1.5$ (continuous line), 
$\alpha=2$ (dashed line), and $\alpha=2.5$ (dot-dashed line)}
\label{TPAGBlifetime_ZsolZlmc}
\end{figure}

Fig.\,\ref{TPAGBlifetime_dmZlmc} shows the evolution of the TP-AGB
lifetime as a function of initial mass for the metallicity of the LMC,
using different mass loss prescriptions as indicated in the
caption. The plot shows that, for any given mass-loss
prescription, the TP-AGB phase is an increasing function of stellar mass
in the low mass range (M$_{i}\,\leq$\,2.5\,M$_{\odot}$),
while for higher masses the trend is the opposite. The figure clearly shows how
sensitive the TP-AGB lifetimes are to the adopted mass-loss prescription.

Note that while the initial-final mass relation does not show a strong
dependence on calibrated mass-loss prescriptions, the lifetime of the TP-AGB
phase does. The initial-final mass relation does not tell us how
how rapidly material is burnt during the TP-AGB phase. 
These considerations are of great interest for population synthesis models of
resolved stellar populations (Mouhcine \& Lan\c{c}on 2002).

Fig.\,\ref{TPAGBlifetime_ZsolZlmc} compares the evolution of the TP-AGB phase
lifetime as a function of initial mass for solar and LMC metallicities.
The plot shows that the total TP-AGB lifetime increases with
decreasing metallicity. This is due to the fact that low-metallicity stars
have globally warmer asymptotic giant branches and need to evolve to 
higher luminosities in order to reach the superwind regime and to end 
their TP-AGB lives. Raising the value of the mixing length parameter
increases the relative lifetime of the TP-AGB phase.
This is related to the fact that increasing $\alpha$ produces
warmer effective temperatures, making the
phase lifetime longer. However, the quantitative
dependence on $\alpha$ remains small.  The derived average
lifetime of luminous, massive TP-AGB stars (M$_{init}\ga\,3.5-4$M$_{\odot}$, 
$-7\la\,M_{bol}\la-6$, mean period of variability
500-900 days) is of the order of a few 
10$^{5}$\,yr, which is consistent with the observational constraints 
(Reid et al. 1990, Hughes \& Wood 1990; see Sect.\,\ref{pulse_LPV}). 
At both metallicities, 
maximal TP-AGB lifetime is obtained for stars with initial masses 
of 2\,--\,2.5\,M$_{\odot}$, which evolve off the main sequence several
10$^8$\,yrs after they were born.

\begin{figure}[h]
\includegraphics[clip=,angle=0,width=0.49\textwidth]{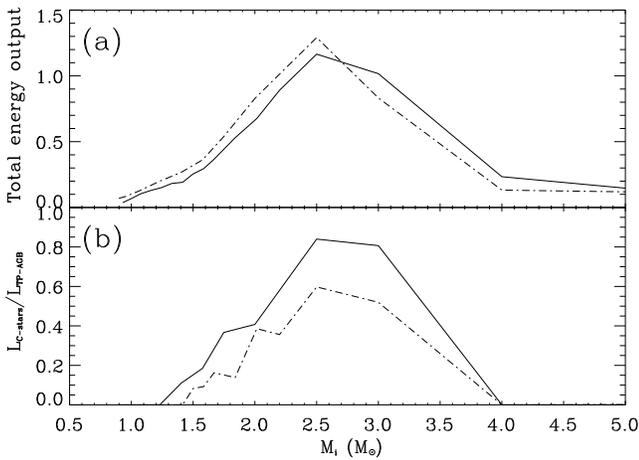}
\caption[]{(a) Total amount of energy (in units of $10^{4}\,$L$_{\odot}\,$Myr) 
produced during the TP-AGB phase for LMC (solid line) and solar (dashed line) 
metallicities, as predicted by our standard models ($\eta=0.1$,
$\alpha=2$, $\lambda=0.75$).
(b) Corresponding fraction of the total energy emitted by the TP-AGB stars 
while they are carbon-rich stars.}
\label{fuel}
\end{figure}

Related to the total lifetime of the TP-AGB phase is the total amount 
of energy emitted during this phase, which is closely related to the 
so-called fuel consumption.
Fig.\,\ref{fuel} shows the total TP-AGB energy output and the fraction 
of this output produced by carbon stars (Sect.\,\ref{cstars.sec}), 
for the two metallicities investigated. 

\subsubsection{Pulsating stars on the AGB}
\label{pulse_LPV}
While evolving along the AGB phase, stars may, for at least part 
of their lifetime, undergo large amplitude, long period pulsations
(period of $100-1000\,$days). 
This evolutionary feature applies to the two classes of LPVs, namely 
the optically visible Miras and semi-regular variables, with low or 
moderate mass-loss rates, and the dust-enshrouded OH/IR or "C/IR" 
stars (carbon-rich infrared sources), with high mass-loss rates
(Wood et al. 1992, 1998; Wood 1998). 

\begin{figure}[ht]
\includegraphics[clip=,angle=0,width=9.cm,height=8.cm]{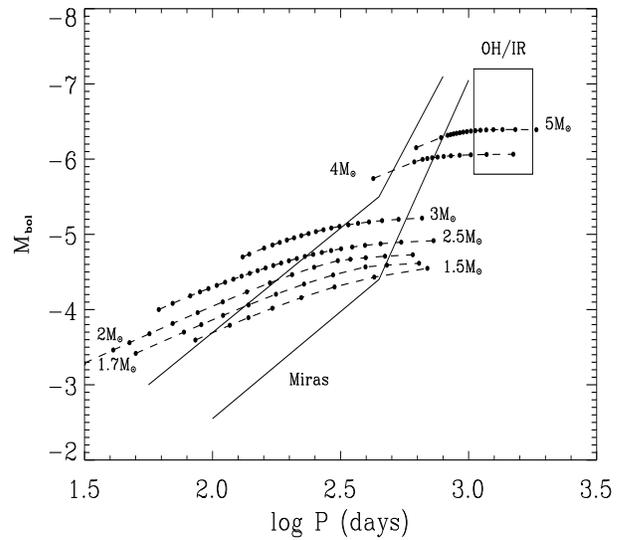}
\caption[]{The fundamental mode period\,--\,luminosity relations for 
stars with initial chemical composition [Z=0.008, Y=0.25] and
our standard model parameters. Initial masses are indicated.
The observed P-L relation for optically visible LPVs in 
the LMC is delineated by the two full lines from Hughes \& Wood (1990).
Also shown is the observed location of dust-enshrouded oxygen-rich AGB 
(OH/IR) stars from Wood et al. (1992). 
The evolutionary points plotted here correspond to the maximum luminosity 
during the quiescent hydrogen burning phase of the helium flash cycles.}
\label{P_L}
\end{figure}

Coupled with a period-mass-radius relation (Wood 1990), 
the present model predicts the period range that LPVs on the
TP-AGB may have. Fig. \ref{P_L} illustrates the results at the
metallicity of the LMC. For each mass,
it shows the bolometric magnitude M$_{bol}$ as a function 
of pulsation period, for the evolutionary points just prior to the 
occurrence of each thermal pulse assuming that LPV pulsation occurs 
during the entire TP-AGB phase. For all masses the first point shown 
corresponds to the first thermal pulse on the TP-AGB, while successive 
points on the tracks are separated by one interpulse period. We assumed 
that LPVs are fundamental mode pulsators, which is now considered 
appropriate for Miras (Wood et al. 1999). When exactly LPV pulsation
starts on the TP-AGB, and when the transitions
from overtone pulsation to lower orders and finally to fundamental
mode pulsation occur, is presently unknown because the
necessary non-linear pulsation models of these complex cool
stars are rare and uncertain (Barth\`es \& Luri 2001 and references
therein). Initially, the stars evolve up the AGB increasing in luminosity and 
period until the onset of the superwind regime, at which point the 
period quickly grows to higher values. The acceleration to longer 
period is more dramatic for massive stars. This is  
because there is more mass to lose, and because the period strongly 
depends on both the mass and the effective temperature. Among massive 
stars (M$_{init}\ga\,3\,$M$_{\odot}$), the final
period is an increasing fonction of initial mass.

We show in Fig.\,\ref{P_L} the zone occupied by optically visible LPVs 
(Hughes \& Wood 1990) and dust-enshrouded stars (Wood et al. 1992), 
as observed in the LMC. The agreement with the predictions 
is satisfactory. We note that for 
M$_{init}\,\la\,2.5-3\,$M$_{\odot}$ the tracks are essentially 
located in the region occupied by the optically visible variable 
stars, before the onset of a brief superwind phase. More massive 
tracks evolve rapidly from the location of optically visible variables
to the location of stars in the superwind regime. This may indicate 
that only relatively massive AGB stars 
(i.e., M$_{init}\ge\,3.5-4\,$M$_{\odot}$) are able to evolve as 
heavily dust-enshrouded objects. The low luminosity zone of the 
observed Period-Magnitude relation where there are no evolutionary 
tracks can be explained by low-mass stars in the luminosity dip 
that follows a thermal pulse.

\subsubsection{Carbon stars}
\label{cstars.sec}

One of the main goals of this paper is to introduce carbon stars into
spectrophotometric models in a self-consistent way.
Using our TP-AGB synthetic evolution models, we thus 
investigate the formation of carbon stars. Here we present some properties of 
these carbon stars that allow us to understand their effects on integrated 
near-infrared properties and the sensitivity of the latter to the metallicity. 
The distribution and the effects of carbon stars on the stellar population 
properties depends on the lifetime of stars in the carbon-rich phase. 
This lifetime depends primarily on the initial mass 
and metallicity.

\begin{figure}[h]
\includegraphics[clip=,angle=0,width=9.cm,height=8.cm]{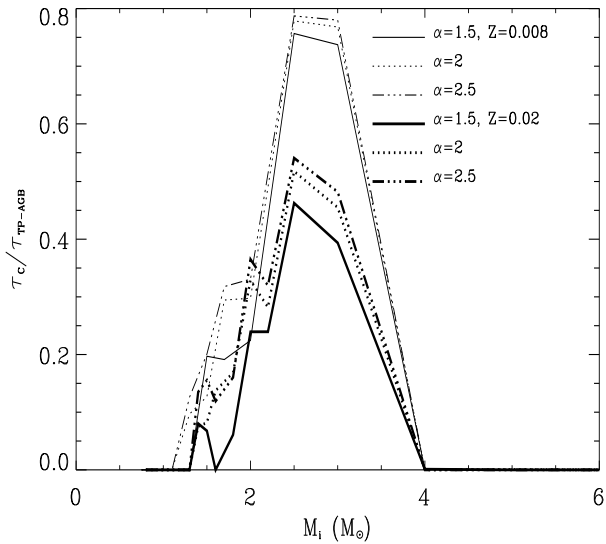}
\caption[]{Fraction of the TP-AGB lifetime spent as a carbon star, for solar
metallicity and LMC metallicity, assuming different mixing length parameters
as indicated. The mass loss prescription from Bl\"ocker (1995)
was used with $\eta\,=\,0.1$.}
\label{tauC_calib}
\end{figure}

In Fig.\,\ref{tauC_calib} we plot the theoretical fraction of the total TP-AGB
lifetime that a star spends as a carbon star, versus the initial mass, for 
the two metallicities considered in this paper and different choices of 
the mixing length parameter. The plot shows clearly that the production 
of carbon stars is a function of both the metallicity and the initial stellar 
mass. In an instantaneous burst scenario, the carbon star subpopulation 
will thus depend on the age of its parent population. 
The striking features are: (i) at a given metallicity, the relative lifetime 
of the carbon-rich phase is an increasing function of stellar mass in the low 
mass range (M$\,\la\,2.5-3\,$M$_{\odot}$), while for higher masses the trend 
is reversed, and (ii) at a given initial mass the relative lifetime
of the carbon-rich phase is significantly larger for models with lower 
metallicity. This behaviour is consistent with observational findings
(see Mouhcine \& Lan\c{c}on 2002 for references and discussion).
In the initial mass range most favourable to the formation of carbon stars,
a rise of the value of the mixing length parameter increases the TP-AGB 
lifetime without significantly changing the time it takes a star to 
become carbon-rich. The carbon star lifetime relative to the TP-AGB 
lifetime thus increases. However, the increase remains small if compared
to the effect of metallicity.

It is noteworthy that $\tau_{\rm C}/\tau_{\rm TP-AGB}$ is largest 
for initial masses that also have the longest TP-AGB lifetime. This 
is clearly demonstrated by the effect of the mixing length parameter
$\alpha$ on the evolution of $\tau_{\rm C}/\tau_{\rm TP-AGB}$. 
Increasing $\alpha$, increases $\tau_{\rm C}/\tau_{\rm TP-AGB}$  
similarily to its effect on $\tau_{\rm TP-AGB}$. The longest relative 
lifetimes as carbon stars indeed correspond to the longest absolute 
lifetime of the carbon-rich phase. 

Another way of illustrating the role of carbon stars is to look 
at the integrated bolometric luminosity emitted during this phase. 
Figure \ref{fuel}\,b shows the predicted fraction of the total amount 
of TP-AGB fuel burnt during the carbon-rich phase versus the initial 
mass of the stars. The features of  these curves
are qualitatively similar to those noted for the time 
fraction that TP-AGB stars spend as carbon-rich objects. 
This directly follows from the fact that the integrated
energy emitted by any evolutionary phase scales with the lifetime
and with the average luminosity of this phase. 
Along an evolutionary track, the average luminosity of the carbon
stars is only slightly higher than that of the oxygen-rich TP-AGB stars, 
and hence the ratio of energy outputs is only slightly 
higher than $\tau_{C}/\tau_{TP-AGB}$. The figure foreshadows that 
the contribution of carbon stars to the integrated properties of stellar 
populations will be maximum when the turn-off mass is 
M$_{TO}\,\approx\,2.5\,$M$_{\odot}$.

\begin{figure}[ht]
\includegraphics[clip=,angle=0,width=9.cm,height=8.cm]{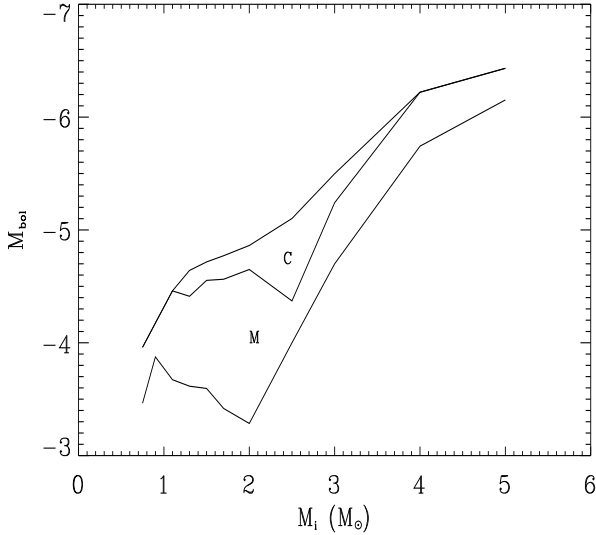}
\caption[]{Bolometric magnitude of important transitions
for TP-AGB stars with initial metallicity Z=0.008. Shown are
the transitions from the E-AGB to the TP-AGB (lowest line), 
the transition from spectral type M to spectral type C (intermediate
line), and the end of the TP-AGB (top line), as functions of initial mass. }
\label{Mi_MbolZlmc}
\end{figure}

\begin{figure}[ht]
\includegraphics[clip=,angle=0,width=9.cm,height=8.cm]{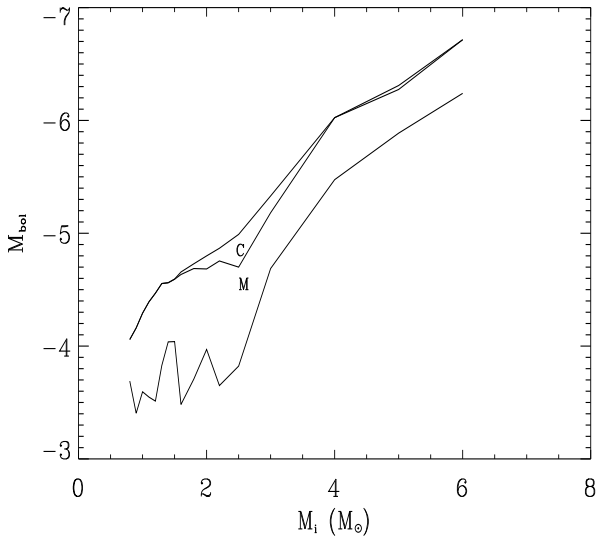}
\caption[]{Same as Fig.\,\ref{Mi_MbolZlmc}, but for
the initial metallicity Z=0.02. }
\label{Mi_MbolZsol}
\end{figure}

Figures \ref{Mi_MbolZlmc} and \ref{Mi_MbolZsol} show the bolometric 
magnitude M$_{bol}$ at which stars with various initial masses 
(M$_{init}$) reach the TP-AGB, become carbon stars, and leave the TP-AGB.
The three boundary lines delimit two areas in the M$_{bol}$--M$_{init}$ 
diagram: the region marked ``M" corresponds to the locus of TP-AGB stars 
of spectral type M (or K; with $C/O\,<\,1$); the region marked ``C" 
represents the domain occupied by the carbon stars (defined as stars 
having $C/O\,\ge\,1$). Note that 
the luminosities plotted are quiescent luminosities, and that the
flash-driven light curves will temporarily move stars of a given
type outside the delineated areas. 
Two features arise from these plots.
The first one is that the bolometric magnitude at the tip of the TP-AGB phase 
is strongly dependent on the metallicity as well as on the initial mass of 
the star: the lower the metallicity, the more luminous is the tip of the AGB 
for a given initial mass (see also Marigo et al. 1996\,a). 
One needs to keep this property in mind when attempting to derive 
constraints on the age of stellar populations: the high end
of the luminosity function of TP-AGB stars is sensitive to metallicity.
The second feature concerns the dips in the M-to-C transition curves
around M$_{init}=2.5\,$M$_{\odot}$. The dip is more 
pronounced at LMC metallicity than at solar metallicity. Note that the 
prefered initial mass range for the formation of carbon stars is 
similar for solar and LMC metallicities in the M$_{bol}$--M$_{init}$
diagram.

\subsection{The stellar spectral library}
\label{spec_lib}

Theoretical and empirical stellar libraries are becoming more and more reliable, 
and have now included most evolutionary phases of stellar evolution. 
However, none of the stellar libraries commonly used for the evolutionary
synthesis of stellar population spectra (e.g. Lejeune et al. 1998;
Pickles 1998) accounts for the specific spectral properties of luminous 
AGB stars. Previous attempts to model intermediate-age populations use static
giant star spectra to model the spectral energy distribution even
of LPVs on the asymptotic giant branch. However, the
library of spectra of luminous red stars of Lan\c{c}on \& Wood (2000)
has confirmed that those stars have spectra quite different from those 
of static M giants. For instance, their near-IR spectra display  stronger
molecular absorption bands than static giant star spectra, as 
a consequence of pulsations that make these stars more extended and 
cooler than static ones (Bessell et al. 1989).  
In addition, no previous evolutionary synthesis predictions explicitly 
include spectra of carbon stars (only broad band colours of these objects
have sometimes been used). In this paper, we use the spectra
of Lan\c{c}on \& Wood  (2000) for static luminous red stars, the 
average spectra of Paper~II for oxygen rich and carbon rich LPVs,
and the library of Lejeune et al. (1998) in all other cases.
We assume that early-AGB stars are static, and TP-AGB stars variable.
The reader is referred to Paper~II for a discussion.

\section{Evolution of integrated properties} 
\label{int_prop}

In this section, we use the TP-AGB evolutionary tracks described 
above to investigate the near-IR properties of single-burst populations. 
As discussed, we now have at our disposal the following essential elements
of such calculations and of their interpretation:
\begin{enumerate}
\item the location of TP-AGB stars in the HR diagram as a function of initial 
      mass and initial metallicity;
\item the total lifetime of the TP-AGB phase;
\item the age marking the transition from M star class ($C/O\,<\,1$) 
to carbon star class ($C/O\,\ge\,1$);
\item a spectral library including oxygen-rich and carbon-rich TP-AGB stars.
\end{enumerate}
In the following, we will present the contribution of AGB
stars to the integrated light and the evolution of integrated optical/near-IR 
colours. In all our synthetic populations, stars are distributed at birth
according to the initial mass function (IMF) of Salpeter (1955; 
$\phi(m)\,\propto\,m^{-2.35}$).

\subsection{Fractional contribution of AGB stars to the total luminosity.}

\begin{figure}
\includegraphics[clip=,angle=90,width=9.cm,height=8.cm]{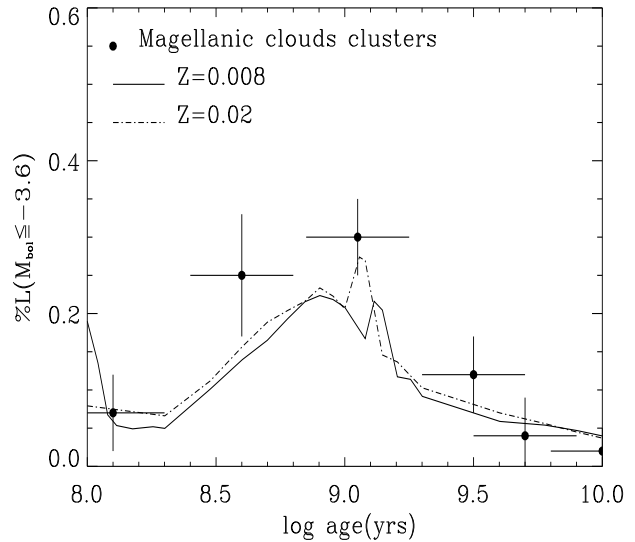}
\caption[]{The predicted fractional bolometric luminosity of single stellar 
populations that arises from AGB stars, as a function of time, for the 
metallicities indicated.
LMC cluster data are adapted from Frogel et al. (1990)
using the age calibration of Girardi et al. (1995). Error bars indicate 
1\,$\sigma$ uncertainties due to small number statistics, but neglect 
membership uncertainties in the samples.}
\label{AGB_contrib}
\end{figure}

Analyzing the contribution of AGB stars to bolometric light is instructive 
in the sense that it tells us when those stars are important 
in the energy budget of a population. The temporal evolution of the 
contribution of the AGB stars to the bolometric light essentially
reflects the evolution of the AGB lifetime as function of the initial mass 
(i.e. Fig.\,\ref{TPAGBlifetime_dmZlmc}).
This is illustrated in Fig.\,\ref{AGB_contrib}. Instead of
the contribution of the TP-AGB stars {\em stricto sensu}, 
the contribution of stars brighter than M$_{bol}=-3.6$  is shown,
in order to allow for direct comparison with existing observations 
(Figs.\,\ref{Mi_MbolZlmc} and \ref{Mi_MbolZsol} show that this limit is 
close to the actual TP-AGB limit at most relevant ages).

The data represent the fractional bolometric luminosities 
accounted for by such stars in clusters of various ages in
the Magellanic Clouds (Frogel et al. 1990).
AGB stars are intrinsically rare objects. In intermediate-age 
star clusters such as those of the Magellanic Clouds,
with masses of the order of 10$^5$\,M$_{\odot}$ or smaller,
the number of luminous AGB stars varies between a small 
handful and about 20. The number and luminosity of AGB stars
are subject to stochastic fluctuations from one cluster to another 
one of similar age, and therefore so are the AGB contributions to the 
integrated light (Ferraro et al. 1995, Girardi et al. 1995, 
Santos \& Frogel 1997, Lan\c{c}on \& Mouhcine 2000). 
To limit this effect, the 39 clusters of the Frogel et al. (1990) 
sample have been grouped into age bins. 
The error bars on the luminosity contribution in Fig.\,\ref{AGB_contrib}
represent the r.m.s. dispersion of these data at $1\,\sigma$, but 
do not include membership uncertainties.
Using the clusters common to the two samples, we determined the linear 
transformation from the SWB class given by Frogel et al. (1990; see 
Searle, Wilkinson \& Bagnuolo 1980) to the S parameter redefined
by Girardi et al. (1995). Then the cluster ages were assigned using the 
relation of Girardi et al. (1995) between the S parameter
and age: $\log (t/yrs)\,=\,0.0733\times\,S+6.227$. 
This relation is based on the same evolutionary models 
as ours (up to the TP-AGB; the latter however has insignificant
impact on the optical colours characterizing SWB classes).

Figure\,\ref{AGB_contrib} shows that our model predictions 
reproduce the variations of the contribution of the AGB stars
with age. The contribution of AGB stars increases gradually from a few percent 
at 0.1-0.2\,Gyr (between SWB type 3 and 5) up to $\sim\,25\%$
at 0.5-1\,Gyr.
This behaviour is due to the effect of envelope burning on massive AGB stars, 
which reduces their lifetime and hence their total emitted light. 
In models without envelope burning, the maximum contribution
of AGB stars occurs at younger ages.  The peak in the AGB contribution 
is located at the age when the turn-off stars have the 
largest TP-AGB lifetime. Our peak values are compatible
with the data in view of the many empirical uncertainties.
Nevertheless, they lie slightly below the observed mean: the important effects 
of AGB stars on the integrated spectrum, discussed in following
sections, are not overestimates.
For populations older than $\sim$\,1\,Gyr, the contribution of AGB stars 
decreases with increasing age as a consequence of shorter lifetimes
and smaller individual luminosities. 
Note that TP-AGB stars dominate the contribution 
of AGB stars to the bolometric light, and account for more than 80\% of the 
total AGB luminosity at $\sim\,1\,$Gyr.

Fig\,\ref{AGB_contrib} shows that the contribution of bright AGB stars 
to the bolometric light is not very sensitive to the metallicity in the
range explored here. This is mainly because the core mass and the
envelope mass at the start of the TP-AGB phase do not strongly depend on 
initial stellar metallicity.
The sharp and short increase of the contribution of AGB stars  
that takes place at $\log(age)\,\approx\,9.15$ 
is due to the first emergence of AGB stars that have degenerate helium 
cores during their RGB phase (see Sect.\,\ref{colevol.sec} and 
Fig.\,5 of Girardi \& Bertelli 1998). 

\begin{figure}
\includegraphics[clip=,angle=90,width=9.cm,height=8.cm]{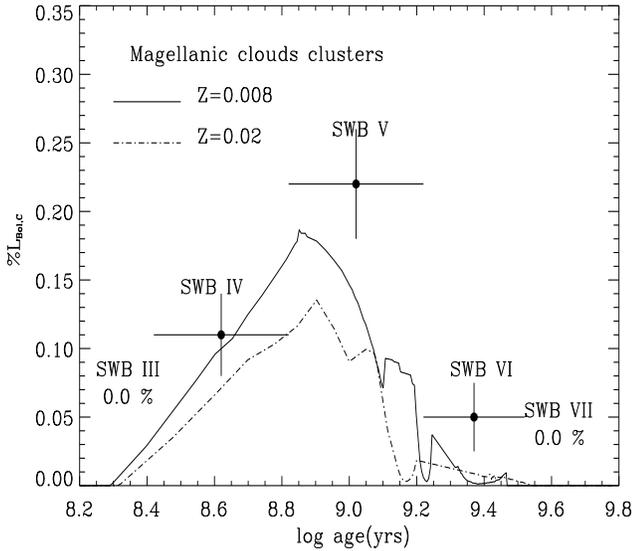}
\caption[]{The fractional bolometric luminosity of single stellar populations 
that arises from carbon stars, as a function of time, for LMC 
(continuous line) and solar metallicity (dashed-dotted line). 
LMC cluster data are adapted from Frogel et al. (1990),
as in Fig.\,\ref{AGB_contrib}.
The lower the metallicity, the higher the contribution of carbon stars.
The maximum contribution corresponds to the age 
at which stars with the turn-off mass have the longest carbon rich phase.} 
\label{C_contrib}
\end{figure}

Figure \ref{C_contrib} shows the evolution of the contribution of 
carbon stars to the integrated bolometric luminosity of a single-burst 
population for Z=0.008 and Z=0.02. 
Overplotted are the empirical fractional bolometric luminosities of 
carbon stars as observed in Magellanic Clouds clusters (Frogel et al. 1990). 
The horizontal error bars refer to the age interval 
covered by the considered SWB types. 
Similarly to the total AGB contribution, the fractional contribution 
of carbon stars is shaped by the evolution of the lifetime in the carbon-rich 
phase as a function of initial mass and initial metallicity 
(Fig. \ref{tauC_calib}).   
Comparison with Fig.\,\ref{AGB_contrib} shows that, when present, carbon stars 
account for a large fraction of the bolometric luminosity from the AGB stars. 
The models show that carbon stars are present only for stellar 
populations older than $\sim\,0.2\,$Gyr, between SWB type 3 and 4 
(corresponding to a turn-off mass M$_{TO}\,\la\,4\,$M$_{\odot}$ with 
very weak dependence on initial metallicity). This is consistent with the 
observational constraints from the Magellanic Cloud clusters. The figure 
shows also that carbon stars appear earlier in metal-poor systems than in  
metal-rich systems, reflecting both the high efficiency of carbon star
formation at low metallicity and the dependence of the total lifetime 
on metallicity. Then, the contribution of carbon stars increases with age, 
with a maximum of $\sim\,20\%$ at age of $\sim\,0.8\,$Gyr. As pointed out for 
the contribution of luminous AGB stars, 
at this age the turn-off stars have the largest carbon-rich lifetime. 
For older systems (age$\,\ga\,0.8\,$Gyr), 
the contribution of carbon stars decreases with age.

The plot shows that, as expected, the fractional contribution of 
carbon stars depends strongly on the metallicity. Lower initial metallicity 
leads to higher contributions of carbon stars to the bolometric light. 
However, the age when the contribution of carbon stars is maximum is 
relatively independent of the initial stellar population 
metallicity, reflecting that the prefered mass range for the formation of 
carbon stars is the same for both metallicities.

Comparison of the model predictions with the observed data and the 
conclusions one may draw from this exercise again require caution.  
The carbon star samples 
are small and the dispersion in the data large. Again, the plotted 
uncertainties are r.m.s deviations that do not account for the many 
uncertain membership (see Table\,1 of Frogel et al. 1990). As an example, 
the LMC clusters NGC 2209 and NGC 269 have the same SWB classification 
(i.e. the same age), but show completely different contributions from 
carbon stars. Carbon stars are responsible for 58\% of the bolometric 
light of the first cluster, while they do not contribute at all to the 
integrated light of the second cluster.
More powerful constraints on the formation of carbon stars, and their 
contribution to integrated light of stellar populations, can be derived 
using the observed statistics of late-type stellar contents of the fields of
resolved galaxies in the local universe, for which the problem of the 
small number statistics of bright carbon stars is avoided. The evolution
of the mean bolometric magnitude of carbon stars in the Local Group  
galaxies is of particular interest for our purpose. However, in
order to use galaxy field stars
one needs to disentangle the stellar evolution effects from 
those of the star formation and the chemical evolution of galaxies. 
Our results for Local Group galaxies are discussed elsewhere 
(Mouhcine \& Lan\c{c}on, 2002).

In summary, the predicted shape of the temporal evolution 
of carbon star contributions to the bolometric light 
is consistent with observations in clusters and constraints
from galaxy fields.
The age interval during which carbon stars are important 
contributors to the bolometric light is well matched. 
This is taken as evidence that the evolution of the carbon-rich phase 
lifetime as a function of initial mass and metallicity is satisfactorily 
determined by our models.

\subsection{The contributions of AGB stars to near-IR light}

In addition to the above,
a fundamental and global test for our single 
stellar population models and for the underlying stellar evolution models, 
is the comparison of the predicted luminosity contributions to different 
near-IR bands with measurements in clusters. 
Analyzing these contributions is important because it provides additional 
information on the spectral features to be expected in the integrated spectrum 
in various wavebands.

\begin{figure}
\includegraphics[clip=,angle=90,width=9.cm,height=8.cm]{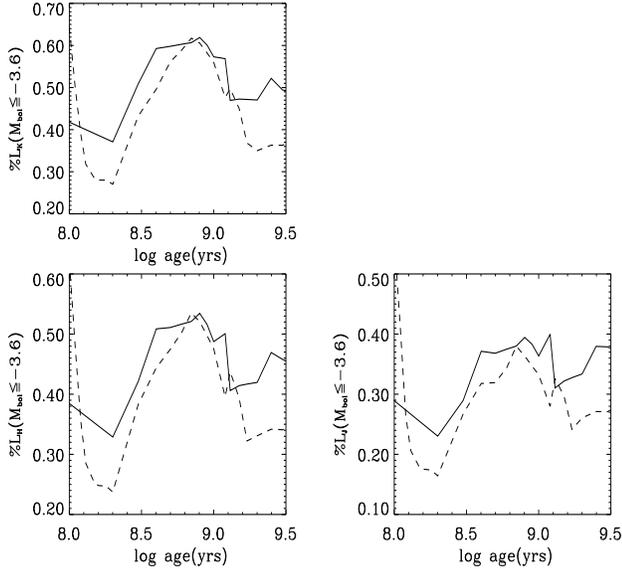}
\caption[]{The fractional near-IR luminosities of single stellar populations
that arises from luminous AGB stars as a function of time for Z=0.02 
(continuous line) and Z=0.008 (dashed line). The redder the pass-band and 
the higher the metallicity, the higher is the contribution of TP-AGB stars.}
\label{AGB_NIR_contrib}
\end{figure}

In Fig. \ref{AGB_NIR_contrib} we show, for the two metallicities 
considered here, the evolving contributions of luminous 
AGB stars to the integrated light in the J, H and K wavebands. 
The plot shows that the evolution of the AGB contributions to near-IR light 
is qualitatively similar to that of their contribution to bolometric light, 
and that it is controlled again mainly by the evolution of the AGB lifetime with
initial mass.  The contribution of AGB stars to integrated 
near-IR light evolves rapidly, for LMC metallicity for example, from a few 
percent at $\sim\,0.1\,$Gyr to being the 
dominant source of the infrared light at ages of $\,\sim\,0.6-0.7\,$Gyr, 
where the contributions vary from $\sim\,40\%$ in the J band to 
$\sim\,65\%$ in the K band. The plot shows that the contribution of AGB 
stars increases with increasing metallicity and increasing wavelength. 
This is because (i) AGB stars are the coolest stars of the population 
and radiate preferentially in the infrared, and (ii) the 
average temperature of AGB stars is sensitive to metallicity (higher 
metallicity lowers the effective temperature of the evolutionary tracks).

\begin{figure}
\includegraphics[clip=,angle=90,width=9.cm,height=8.cm]{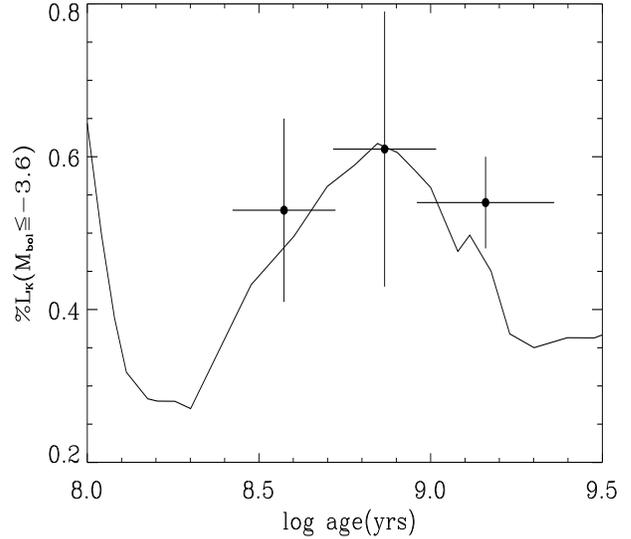}
\caption[]{The AGB contribution to the total K band luminosity compared with
globular cluster data from Ferraro et al. (1995). The solid line represents the
predicted contribution of AGB stars to the K band luminosity for single stellar
population of LMC metallicity. Error bars indicate the r.m.s. dispersion
in the data.}
\label{AGB_Kband}
\end{figure}

In Fig.\,\ref{AGB_Kband} we show a comparison between the temporal 
evolution of the AGB contribution to the integrated K band luminosity, 
as theoretically predicted for Z=0.008 and as observed in a sample of 
intermediate-age LMC clusters (Ferraro et al. 1995). 
Unfortunately no observational constraints are available on the contribution 
of AGB stars to the J band nor the H band luminosities. Globular cluster 
ages and error bars are as noted above for the contribution of AGB stars 
to the bolometric light. The plot shows a good agreement between the observed 
and the predicted K band contributions of the AGB stars.
The agreement suggests that the combination of the predicted position of 
TP-AGB stars on the theoretical HR diagram and their total lifetime
is adequately determined by our stellar evolution models.

\subsection{The time evolution of integrated colours}
\label{colevol.sec}

We now discuss how the integrated colours evolve in time.
We recall that for single-burst populations older than 0.1\,Gyr, the 
near-IR colours depend weakly on the assumed IMF (Maraston 1998). 

\begin{figure*}
\centering
\includegraphics[clip=,angle=0,width=14.cm,height=10.cm]{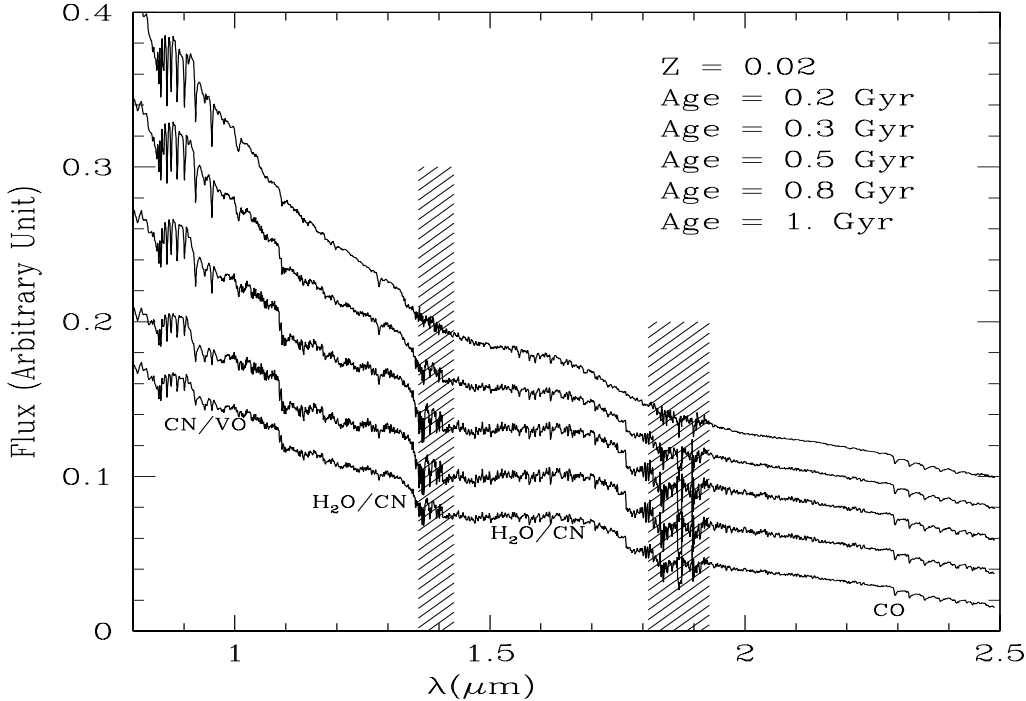}
\caption[]{Evolution of the spectra of instantaneous
burst stellar populations as predicted for Z=0.02 for the ages
given. Main molecular contributors to the features that characterize the
NIR spectra are indicated.
They characterize intermediate-age populations
and may be used to identify AGB populations.
Note particularly the evolution of the H band shape as function of age and
the break-like feature at $1.1\,\mu$m.
Shaded areas are those that could not be corrected for telluric
absorption in a satisfactory way.}
\label{syn_spectra}
\end{figure*}

Figure~\ref{syn_spectra} illustrates the spectral evolution of a solar 
metallicity, instantaneous burst population between 0.2 and 1\,Gyr. 
The synthetic spectra are constructed using isochrones assembled
by means of the evolutionary tracks of Sect.~\ref{tracks} 
and the stellar library presented in Sect.~\ref{spec_lib}. 
The figure shows the spectral features that characterize 
the near-IR spectra of intermediate-age populations. The time evolution 
of the H band spectral energy distribution and of the break-like 
feature in the J band clearly show the effect of the dominant AGB population.  
For stellar populations younger than $\sim\,300\,$Myr, the latter
is oxygen-rich, and the near-IR spectra are characterized 
by features that originate from late type M stars. The H band has a 
bump-like curved shape (due to the minimum in the continuous H$^-$ 
opacity and to the wings of water absorption bands), 
and the J band shows VO absorption at $\sim\,1.05\mu$m. 
For older populations, the H band flattens and the $\sim\,1.77\mu$m 
break due to C$_2$ deepens. The water band cut-off around 1.32\,$\mu$m
is progressively hidden in carbon star features and replaced
with a CN bandhead at 1.35\,$\mu$m. Another CN bandhead becomes
conspicuous at $\sim\,1.08\mu$m.
  
\begin{figure}[h]
\centering
\includegraphics[clip=,angle=0,width=10.cm,height=10.cm]{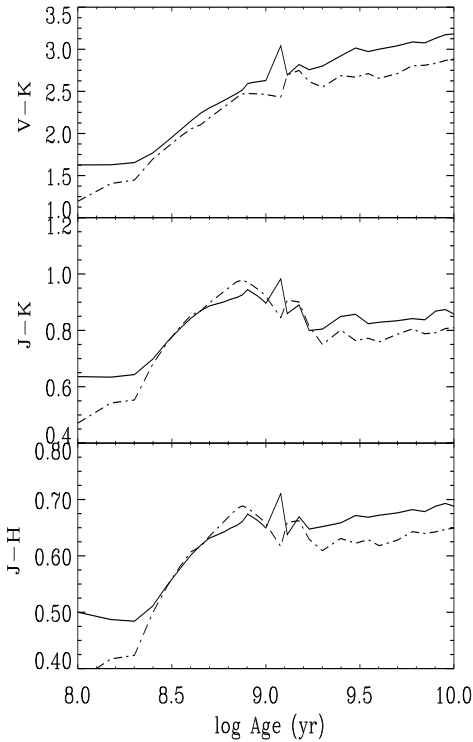}
\caption[]{The time evolution of synthetic broad-band near-IR colours for Z=0.02
(continuous line) and Z=0.008 (dashed-dotted line). The solar models are redder
than the LMC metallicity models apart for intermediate-age populations
where carbon stars are present. For these systems, Z=0.008 models evolve more
rapidly to the red than Z=0.02 models. This behaviour is due to the higher
contribution of carbon stars to the integrated light in metal-poor systems.}
\label{model_Zsol_Zlmc}
\end{figure}

Fig. \ref{model_Zsol_Zlmc} shows the time evolution of the optical/near-IR 
synthetic broad-band colours for the age interval in which AGB stars are 
significant. The integrated VJHK magnitudes of the burst populations 
are obtained by convolving
the spectra with the response functions of standard passbands (Bessell 1990,
Bessell \& Brett 1988). Zero colours are adopted for a model
spectrum of Vega (Dreiling \& Bell 1980, kindly provided by M. Bessell).
The figure illustrates clearly 
that the presence of AGB stars does not {\em immediately} 
translate into a sharp jump of the stellar population colours to the red, 
but that this transition is gradual over several hundred Myr. This predicted 
evolution is in contradiction with the classical view of the evolution of 
optical/near-IR broad-band colours (see e.g. Renzini \& Buzzoni 1986,
Charlot \& Bruzual 1991, Bressan et al. 1994b, Tantalo et al. 1996,
Fioc \& Rocca-Volmerange 1997), as already noted by Girardi \& Bertelli (1998).
The predicted behaviour is due to (i) the severe reduction of the TP-AGB phase 
lifetime of massive stars, and hence the reduction of their contribution to the 
near-IR integrated light (ii) the absence of a clear transition between 
the stars affected by the envelope burning and those that follow the standard 
core mass-luminosity relation. We note that this behaviour is similar for both 
metallicities considered. 
The luminosity excess with respect to original core mass-luminosity relations 
for massive AGB stars causes a temporal smoothing of the colour jumps 
associated with the appearance of AGB stars.

The impact of AGB stars on the integrated properties of intermediate-age 
stellar populations is more evident in purely near-IR colours. 
(V-K) increases as the stellar population ages, 
even at times when the AGB contribution to K band light decreases, 
due to the gradual fading of the turn-off stars that dominate the V-band light 
and the accumulation of large numbers of red giant branch stars that will 
finally dominate the near-IR light. The purely near-IR 
colours, on the other hand, increase up to the age of $\sim\,0.8\,$Gyr, but 
decrease gradually for older systems  up to an age
of $\sim\,1.5-2\,$Gyr. Populations in this age range  are likely
to explain the {\it{IR enhanced clusters}} as defined by Persson et al. (1983). 
The near-IR colours result directly from the large 
contribution of AGB stars to the near-IR light, 
between ages of $\sim\,0.3\,$ and $\sim\,1.5\,$Gyr.
For older systems, the AGB stars are not the main contributors
and the time evolution reflects the well-known accumulation of red 
giants. This picture of the optical/near-IR 
colour evolution is consistent with the 
evolution of these colours observed in the Magellanic Cloud
clusters as a function of SWB type (Fig.\,2 in Persson et al. 1983). 

An interesting feature of Fig.\,\ref{model_Zsol_Zlmc} 
is that the near-IR colours evolve more 
rapidly at Z=0.008 than with Z=0.02 models,
during the age interval where AGB stars are dominant. This follows from
the chemical nature of the predominant AGB populations.
Carbon stars are intrinsically redder than oxygen-rich AGBs in 
colours involving the J band.  As pointed out above, carbon stars 
completely dominate intermediate-age AGB populations at low metallicity,
explaining the very red near-IR colours.

Fig.\,\ref{model_Zsol_Zlmc} shows that there is a brief excursion 
to red colours, just after the age when turn-off stars have an initial 
mass equal to the limit for the development of a degenerate helium
core. As discussed by Girardi \& Bertelli (1998), this transient jump 
is a feature of the evolutionary models, due to the fact that
the core helium burning phase lifetime reduces strongly for stars 
having an initial mass in the vicinity of the transition mass between
stars with and without degenerate helium cores.\\

\begin{figure}
\includegraphics[clip=,angle=90,width=9.cm,height=8.cm]{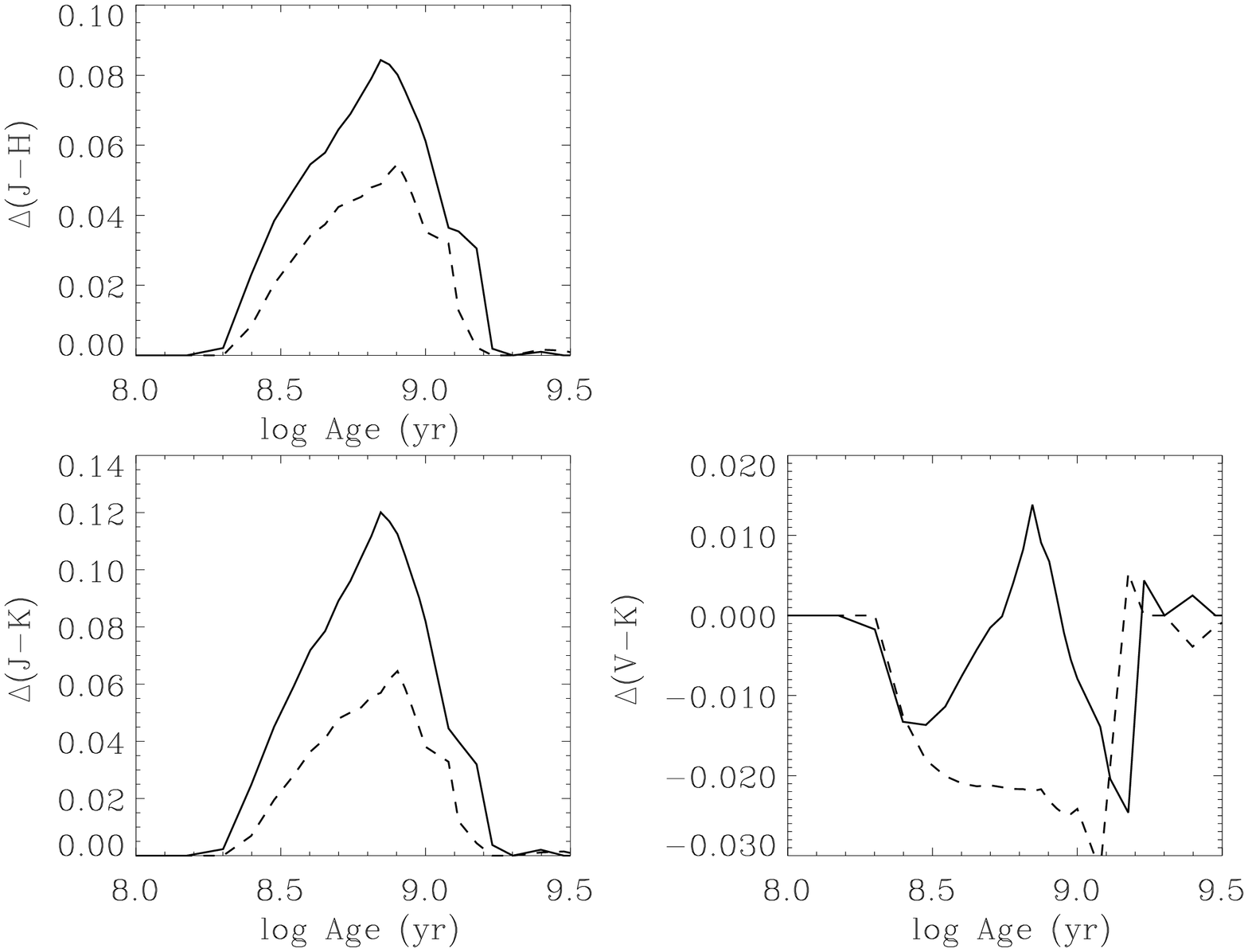}
\caption[]{Differences between the synthetic broad-band colours 
of models that differentiate carbon stars and oxygen-rich stars and models 
that neglect the formation of carbon stars,
for Z=0.008 (continuous line) and Z=0.02 (dashed-line).}
\label{DNIRC_Crich}
\end{figure}

To further illustrate the effect of carbon stars on the integrated light,
we have computed additional model sequences in which the 
formation of carbon stars is neglected, and exclusively spectra of 
oxygen-rich stars are used. Fig.\,\ref{DNIRC_Crich} shows the 
differences between synthetic broad-band colours predicted by 
models that differentiate carbon stars and oxygen-rich stars and 
models that don't. As expected, 
(i) near-IR colours calculated including the two AGB 
subtypes are systematicaly redder than models neglecting the formation 
of carbon stars, (ii) the changes in the colours between both model sets
increase with decreasing metallicity, and (iii) (V-K) is 
not affected significantly by the nature of dominant 
AGB population. The models show that neglecting carbon stars 
has a sizeable effect on the evolution of spectral energy distribution. 
We find that the effects on the J band are significant,
while those on the H and K bands are relatively small. 
Good near-IR photometry including the J band may be able to 
tell us if the AGB populations are dominated by carbon stars or not,
while V-K is not a useful colour for that purpose. 
As an example, the change in (J-K) between both models, for LMC metallicity 
at the age when the AGB contribution is largest (age$\,\approx\,0.8\,$Gyr), 
is $\Delta(J-K)\,\approx\,0.12$, which is easily observable with the 
current instrumentation.\\

\begin{figure*}[!ht]
\centering
\includegraphics[clip=,angle=0,width=15.cm,height=12.cm]{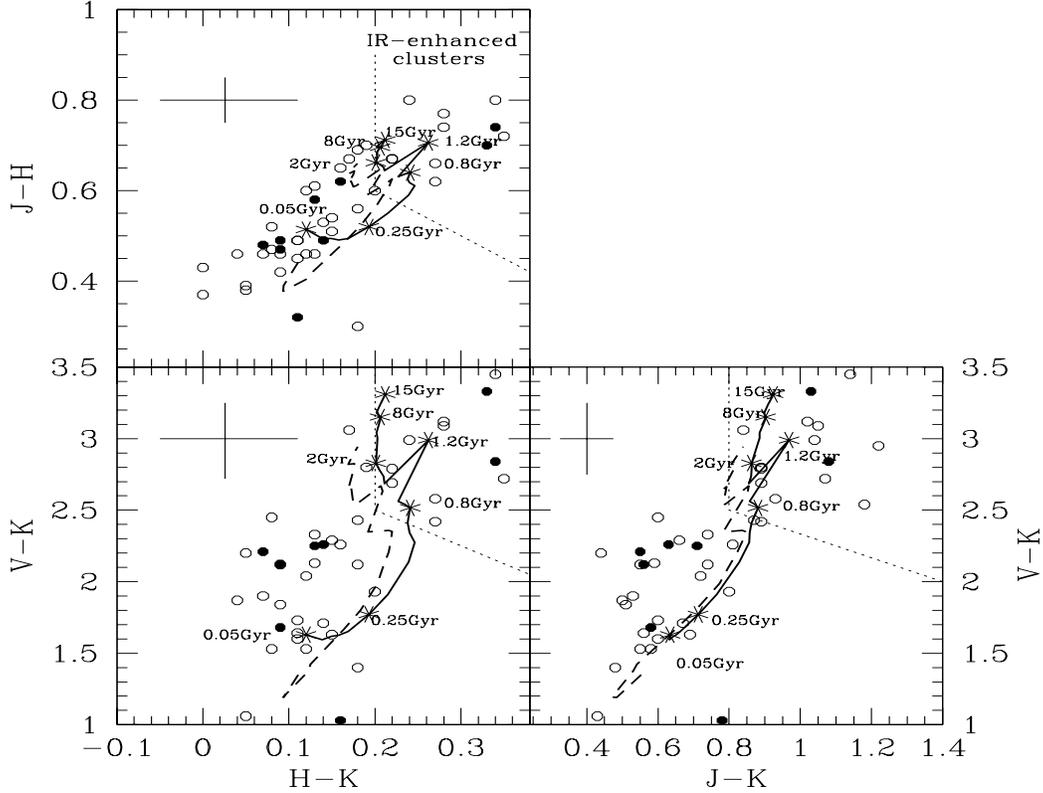}
\caption[]{Comparison of synthetic two-colour diagrams predicted at Z=0.008 
and Z=0.02 (continuous and dashed line, respectively) at ages of 0.1--15\,Gyr 
with observed reddening-corrected colours for star clusters in the LMC 
(open circles) and SMC (filled circles) taken from Persson et al. (1983). 
Typical error bars are indicated. Also shown are age labels along the 
solar model track. The {\em IR enhanced} clusters  of Persson et al. (1983) 
have red colours and a specific range of ages (mean SWB type {\sc iv}). 
Those objects are reproduced by stellar populations dominated by AGB stars.}
\label{model_data}
\end{figure*}

Fig.\,\ref{model_data} shows comparisons between our synthetic colour-colour 
diagram for stellar populations older than 0.1\,Gyr, and observed 
diagrams for a sample of LMC and SMC globular 
clusters with available infrared photometry and SWB types older than 3 
(Persson et al 1983). 
Observational determinations of the age-metallicity relation of Magellanic 
globular clusters (Olszewski et al. 1991, Da Costa \& Hatzidimitriou 
1998), show that old clusters have approximately $-1.5>[Fe/H]>-2.2$, 
while young and intermediate-age ones have $0.0>[Fe/H]>-0.6$.
The significant dispersion of the observed colours, due to the small number 
of bright and red stars, unfortunately limits the power of the comparisons
(Lan\c{c}on \& Mouhcine 2000, Bruzual 2002).
We report in the colour-colour diagrams the location of the IR enhanced 
clusters as defined observationally by Persson et al. (1983). 
The models show that this regions is occupied by stellar populations 
with age $8.5\la\log(t/{\rm yr})\la9.2$ and, hence, dominated by carbon stars. 
This is consistent with the data, showing that (i) almost all of these clusters 
are of intermediate-age, (i.e., in SWB groups IV\,-\,VI), 
and (ii) many members of the IR enhanced cluster group contain 
identified carbon stars, 
which help accounting for their red integrated colours.

\subsection{Searching for AGB stars in post-starbursts}

AGB stars have proven to be powerful tracers of intermediate-age 
stellar populations in nearby resolved galaxies, and hence tools to 
infer constraints on the formation history of these galaxies. 
To retrieve information about relatively recent star formation 
in unresolved galaxies, Lan\c{c}on et al. (1999) suggested 
to use the near-IR signatures of AGB stars present in the integrated spectra. 
They selected a set of intermediate band filters (among the
filters available on the NICMOS camera on board the Hubble Space
Telescope), well suited for measurements of the specific LPV features, 
and appropriate both for the global identification of TP-AGB 
stars and for the determination of their chemical nature.
In this section, we re-evaluate the temporal 
evolution of such a set of photometric molecular indices.

In the following we will use three indices defined as follows: 
H$_{2}$O (2$\mu$m)\,=\,F190N1/F215N2, 
H$_{2}$O (1.64$\mu$m)\,=\,$2\times\,$F190N1/(F164N1+F215N2), and 
H$_{2}$O/C$_{2}$ (1.77$\mu$m)\,=\,F180M2/F171M2. 
The passbands of the selected filters are those of the HST/NICMOS filters. 
All indices are flux ratios (flux in a molecular band divided
by an estimate of the flux outside the molecular band),
expressed in magnitudes, and they take the value 
zero for Vega. 

\begin{figure}
\includegraphics[clip=,angle=0,width=9.cm,height=8.cm]{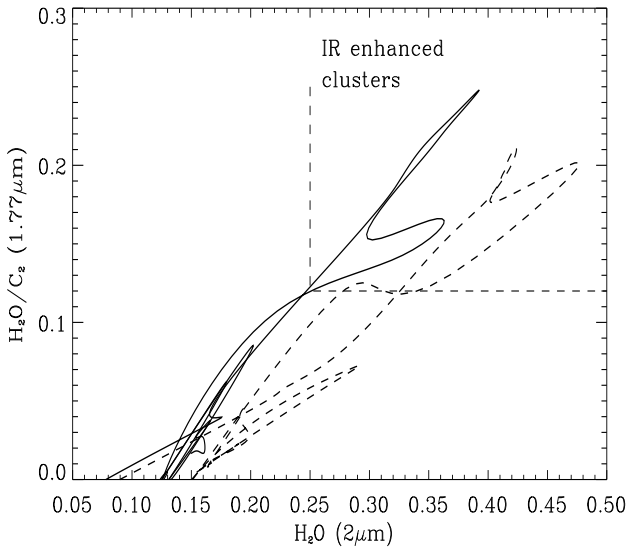}
\caption[]{Synthetic evolution in the prefered two-index diagram.
Solid: Z=0.008; dot-dashed: Z=0.02.
The upper right corner of this diagram is occupied by stellar populations
dominated by AGB stars. They correspond to the IR enhanced clusters
as defined by Persson et al. (1983). }
\label{diagnostic}
\end{figure}

Figure \ref{diagnostic} shows the resulting two-index diagnostic diagram.
High molecular indices unambiguously identify the predominance
of AGB stars in single stellar populations. The plot shows that the
``IR enhanced" stellar populations, defined as above, also
have the most pronounced molecular features. In complex stellar populations,
the identification of intermediate-age post-starburst components
will depend on the extent of dilution in continuum contributions
of either a few red supergiants or many old red giants.
Using two molecular indices jointly should allow us 
to constrain the chemical nature of the dominant AGB subtype, once 
it has been shown that AGB stars are present: here, 
the 1.77\,$\mu$m H$_2$O/C$_2$ index takes systematically higher values for 
populations with higher fractions of carbon stars.

As already mentioned, 
the predictions of near-IR spectrophotometric properties of intermediate-age 
stellar populations requires the use of an effective temperature -- colour 
(or spectrum) calibration for AGB stars. 
We have constructed two grids of models assuming two stellar 
spectral libraries differing in the assumed effective temperature 
scales. The first one is theoretical, and taken from Bessell et al. (1989;
as described in Paper~II).
The authors have considered only static M stars, but with
extended atmospheres meant to be appropriate for AGB stars. 
The second scale was proposed by Feast (1996), based on
angular diameter measurements for late-type stars combining Miras 
and non-Mira M-type stars. A caveat of that scale is the scatter
in the empirical data it is based on. All model result previously
discussed in the present paper are based on the scale 
of Bessell et al. (1989).

\begin{figure}
\includegraphics[clip=,angle=90,width=9.cm,height=8.cm]{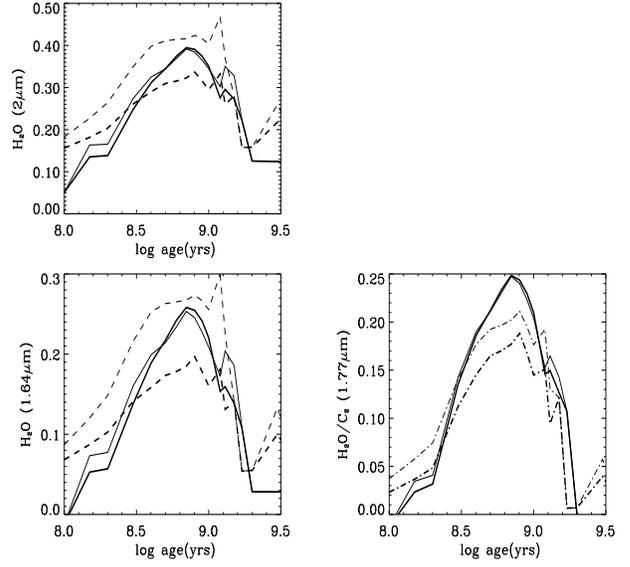}
\caption[]{The time evolution of selected photometric indices 
that identify AGB stars in single post-starburst populations. 
Solid curves: Z=0.008, and dashed line to Z=0.02. Thin lines refer 
to models based on the T$_{\rm eff}$ scale of Bessell et al. (1989),
thick lines to those based on the scale of Feast (1996).}
\label{indices_evol}
\end{figure}

Models that assume the scale of Bessell et al. (1989) predict deeper
molecular bands than those based on the scale of Feast (1996). This 
follows from the steeper slope of Feast's relation, at the low
temperatures relevant to TP-AGB stars: late type stellar spectra 
with deep molecular bands are assigned higher temperatures, and
hence are used much more, in the models with the scale of Bessell et al. 
than in those with Feast's scale. The resulting changes in the molecular
indices are illustrated in Fig.\,\ref{indices_evol}. At solar
metallicity, oxygen-rich TP-AGB stars, although mixed with 
carbon stars, contribute significantly to the light, and the 
temperature scale of the M type spectra is essential. 
Here, our assumption that all TP-AGB stars are variable
also has effects. Using static star spectra rather than LPV
spectra during a reasonable first part of the TP-AGB
would reduce the oxide features slightly.
At Z=0.008 on the other hand, the predominance of carbon stars
is such that changes in the temperature scale of the M type spectra
are negligible.   

The comparison of the new set of theoretical predictions with those
of  Lan\c{c}on et al. (1999) again shows the important
effects of the stellar evolution inputs on the predicted integrated 
properties of intermediate-age stellar populations. 
Lan\c{c}on et al. (1999) used the stellar evolutionary tracks 
provided with the {\sc P\'egase} evolutionary synthesis package 
(Fioc \& Rocca-Volmerange, 1997). 
For early evolutionary phases they are identical to the ones used here, but
their TP-AGB phase is based on the prescriptions of Groenewegen \& de Jong 
(1993) which, in particular, are incomplete for envelope burning. 
The main resulting discrepancy between the two sets of models is 
in the age at which the AGB signatures are the strongest:
without envelope burning, this happens much earlier,
i.e. at $\approx\,0.2\,$Gyr, against $\approx\,0.8-0.1\,$Gyr 
in the present models. An abrupt transition to red colours and
deep molecular features was present around 0.2\,Gyr of age, that has
now turned into a gradual change. In addition, Lan\c{c}on et al. (1999) had
considered the formation of carbon stars quantitatively only
at Z=0.008. 

The more extensive grid that we have now computed
highlights the sensitivity of the molecular indices to
uncertain model parameters. For instance, Fig.\,\ref{indices_evol}
shows that it may depend on the adopted effective temperature scales
whether photometric indices that were designed to
measure H$_2$O absorption  take larger or smaller values in low
metallicity systems than at solar metallicity.
More data is clearly necessary for the calibration of the models.
To avoid the stochastic fluctuations, single stellar
populations more massive than the clusters of the Magellanic Clouds
should be searched for. 
Unfortunately, such objects are rare at observable distances.
One is cluster number 3 in the merger NGC\,7252, a $\sim 300$\,Myr old,
10$^7$\,M$_{\odot}$ candidate. Results from the analysis of 
the near-IR spectrum of this cluster will be presented by Mouhcine
et al. (2002). It would be extremely useful to extend those 
fundamental observations to a sample with a {\em range} of 
intermediate-ages, known for example from optical spectroscopy.

%-----------------
\section{Conclusions}
\label{concl}

In this paper we modelled the spectrophotometric properties of 
intermediate-age stellar populations, focusing on near-infrared wavelengths. 
TP-AGB stars were included using a large grid of synthetic evolutionary
tracks where the effects of envelope burning and the formation of carbon stars
were accounted for in a consistent way, using analytical
expressions to take into account all the relevant physical processes. 
Spectral features of luminous AGB stars were included in the models 
via the new purpose-designed spectral library of Paper~II,
which includes spectra for both oxygen-rich and carbon-rich stars.

We have compared our theoretical predictions with 
observational data from the literature. 
First, our stellar evolutionary tracks are able to reproduce 
the observational properties relative to single AGB stars that may affect 
their contribution to integrated properties of stellar populations (initial 
mass-final mass relation, period-luminosity relation). The predicted 
contributions of bright 
AGB stars and carbon stars to the integrated bolometric luminosity and to
the integrated K band luminosity of a single-burst population are close 
to what is observed in the samples of Magellanic Cloud clusters
by Frogel et al. (1990) and Ferraro et al. (1995). In particular, the
range of ages at which these contributions are maximal is explained. 
The contribution of bright AGB stars to the bolometric 
luminosity is shown not to be a strong function of metallicity, while the 
contribution of carbon stars increases with decreasing metallicity.

The inclusion of envelope burning, the superwind regime at the tip of the 
AGB phase, and the third dredge-up events in stellar evolution 
models are shown to be important in order to match the observational 
constraints. More specifically, we have confirmed that the evolution 
of broad-band optical/near-IR colours are heavily affected by the 
envelope burning from about 0.1\,Gyr to $\sim$1\,Gyr, making the 
colour jump to the red, due to the presence of intermediate-mass stars, 
more gradual and smoothed than suggested in early models. 
We showed also that the inclusion of carbon stars in the models makes 
colours redder than models that consider only oxygen-rich 
stars. The difference between models with and without carbon stars
is an increasing function of metallicity, reflecting the higher 
carbon star formation efficiency at lower metallicity. 
Our instantaneous burst models are able to reproduce well the broad-band VJHK
colours of a sample of Magellanic Cloud clusters
with ages older than 0.1\,Gyr.
The effects of envelope burning and of carbon stars on the properties of 
intermediate-age population are important in describing a 
realistic colour evolution of stellar systems, in particular
the many nearby dwarf galaxies with subsolar metallicities.

We have also re-investigated the evolution of selected narrow-band indices 
(Lan\c{c}on et al. 1999) that are specifically sensitive to the near-IR 
spectra of AGB stars, with the aim of using them to search for post-starburst 
stellar populations. Here again, the effects of envelope burning and
of the formation of carbon stars are large. 
In principle, TP-AGB stars can indeed be identified
in the integrated light of intermediate-age stellar populations; 
and the combination of at least two molecular indices can, at
a range of ages, be used to determine whether oxygen-rich or carbon-rich
stars are predominant, thus providing useful information on 
the metallicity of the environment of which the stars were born. 
However, excellent photometric quality is necessary for such
purposes, and the use of spectroscopy remains preferable (a
broad spectral coverage should be sought). In addition, 
the predictions are rather sensitive to model parameters.
The effective temperature scale of upper AGB stars remains an important
source of uncertainties, and we note that an a posteriori empirical 
calibration of this scale remains essential. A first step in this
direction, based on the most massive star cluster of the 
merger remnant NGC\,7252, will be presented in a companion
paper.

%-----------------
%\begin{acknowledgements}
%.......................
%\end{acknowledgements}

\end{document}